\documentclass{aa}  

\usepackage{natbib,twoopt}
\usepackage[breaklinks=true]{hyperref} 
\bibpunct{(}{)}{;}{a}{}{,}             
\makeatletter
  \newcommandtwoopt{\citeads}[3][][]{\href{http://adsabs.harvard.edu/abs/#3}%
    {\def\hyper@linkstart##1##2{}%
     \let\hyper@linkend\@empty\citealp[#1][#2]{#3}}}
  \newcommandtwoopt{\citepads}[3][][]{\href{http://adsabs.harvard.edu/abs/#3}%
    {\def\hyper@linkstart##1##2{}%
     \let\hyper@linkend\@empty\citep[#1][#2]{#3}}}
  \newcommandtwoopt{\citetads}[3][][]{\href{http://adsabs.harvard.edu/abs/#3}%
    {\def\hyper@linkstart##1##2{}%
     \let\hyper@linkend\@empty\citet[#1][#2]{#3}}}
  \newcommandtwoopt{\citeyearads}[3][][]%
    {\href{http://adsabs.harvard.edu/abs/#3}
    {\def\hyper@linkstart##1##2{}%
     \let\hyper@linkend\@empty\citeyear[#1][#2]{#3}}}
\makeatother
\usepackage{graphicx}
\usepackage{txfonts}
\usepackage{amsmath} 
\usepackage{amssymb}
\usepackage{hyperref}
\hypersetup{
  colorlinks   = true, 
  urlcolor     = blue, 
  linkcolor    = blue, 
  citecolor   = blue 
}
\usepackage{natbib}
\usepackage{orcidlink}
\usepackage{lscape} 
\usepackage{supertabular}
\usepackage{caption}
\usepackage{longtable}
\usepackage{multicol}
\usepackage{booktabs}
\usepackage{supertabular}

\bibpunct{(}{)}{;}{a}{}{,}

\newcommand{\ms}{m\,s$^{-1}$}

\newcommand{\re}{\ensuremath{\mathrm{R_{\oplus}}}}%

\newcommand{\kms}{${\rm km\,s^{-1}}$}
\newcommand{\kmsd}{${\rm km\,s^{-1}\,d^{-1}}$}

\begin{document}

   \title{Homogeneous planet masses I}

   \subtitle{Reanalysis of archival HARPS radial velocities}

   \author{H. L. M. Osborne, \inst{1,2}
          L. D. Nielsen, \inst{1,3}
          V. Van Eylen, \inst{2}
          \and
          O. Barrag\'an \inst{4,5}
          }

   \institute{European Southern Observatory, Karl-Schwarzschild-Str. 2, 85748 Garching bei M\"unchen, Germany
   \and
    Mullard Space Science Laboratory, University College London, Dorking RH5 6NT, UK
    \and
    University Observatory Munich, Ludwig-Maximilians-Universit\"at, Scheinerstr. 1, 81679 Munich, Germany
    \and
    Sub-department of Astrophysics, Department of Physics, University of Oxford, Oxford OX1 3RH, UK
    \and 
    Department of Physics, University of Warwick, Coventry CV4 7AL, UK
    }

   \date{Received September 15, 1996; accepted March 16, 1997}

 
  \abstract{
  Empirical exoplanet mass-radius relations have been used to study the demographics and compositions of small exoplanets for many years. However, the heterogeneous nature of these measurements hinders robust statistical analysis of this population, particularly with regard to the masses of planets. For this reason, we perform a homogeneous and consistent re-analysis of the radial velocity (RV) observations of 87 small exoplanets using publicly available HARPS RV data and the fitting toolkit $\texttt{Pyaneti}$. For the entire sample, we run 12 different models to investigate the impact of modelling choices, including the use of multi-dimensional Gaussian Processes (GPs) to mitigate stellar activity. We find that the way orbital eccentricity is modelled can significantly impact the RV amplitude found in some cases. We also find that the addition of a GP to mitigate stellar activity does impact the RV amplitude found - though if the GP is modelled on activity indicators as well as the RVs the results are more robust. The RV amplitude found for every planet in our sample using all the models is made available for other groups to perform demographics studies. Finally, we provide a list of recommendations for the RV community moving forward.}


   \maketitle
%

\section{Introduction}

While many exoplanets have now been confirmed, significantly fewer have a mass measurement. This means that efforts to characterise the planets, and their planetary systems, in more detail is hindered. Planet mass measurements are important for several reasons: combined with radii measurements they provide a bulk density estimate which can be used to constrain compositions \citep{2014Rice,2023goffo}; the predominant method of finding masses also provides information on system architectures and multiplicity; and a precise mass measurement is essential for atmospheric characterisation \citep{2019ApJ...885L..25B,2023A&A...669A.150D}. Despite this importance, only 18$\%$\footnote{According to the NASA Exoplanet Archive Composite Planet Parameters Table Accessed 31-07-2024} of known small (R $<$ 4 \re) exoplanets have a mass measurement, and even fewer have precise (uncertainty $<$ 20$\%$) mass measurements. To fully understand the exoplanet systems we have detected so far, it is essential that we have precise mass measurements. 

Transit observations show a distinct bi-modality in planet radius distribution, with few planets found around 1.5 to 2 Earth radii \citep{fulton_california-_2017,2018MNRAS.479.4786V}. This radius ‘valley’ separates the smaller super-Earths, which are thought to be rocky in nature, and the larger (2-4 Earth radii) mini-Neptunes, whose composition remains unknown. A variety of models have been proposed to explain the appearance of this radius valley. Some suggest that it is a result of atmospheric loss through either photo-evaporation \citep{owen_evaporation_2017, fulton_california-_2017, 2018MNRAS.479.4786V}, core-powered mass-loss \citep{2018MNRAS.476.2542C,gupta_sculpting_2019,2021MNRAS.504.4634G}, or a combination of these. Others have suggested that the radius valley is a result of a diversity in core composition at formation, with some planets being formed with rocky cores and some with ice/water, see e.g. \cite{luque_density_2022}. Some potential water-world planets are presented in \cite{2024MNRAS.52711138O,2022diamondlowe, cadieux_toi-1452_2022}. To fully understand these possible causes of the radius valley, a good measure of small planet compositions is essential. By characterising small planet masses, we enable their compositions and internal structure to be constrained.

Many current and future exoplanet-focused missions aim to characterise the atmospheres of small planets. Around 30$\%$ of the observing time requested for JWST in Cycle 3 was for the topic of exoplanets and disks, and a new Directors Discretionary Time proposal is focused specifically in finding the atmospheric components of small planets orbiting M type stars. Further ahead, the Plato mission \citep{2014Rauer}, due to launch in 2026, has the aim of characterising the bulk properties (mass, radius, composition) of small planets orbiting bright stars. The Ariel mission will also launch toward the end of the decade, focusing on characterising the atmospheres of around 1000 exoplanets \citep{2018Tinetti}. For all these missions it is essential to not only have precise mass measurements, but also to understand the impact of homogeneous (or in-homogeneous) modelling on the planet masses found.

The vast majority of existing mass measurements come from radial velocity (RV) measurements. This is where high resolution spectrographs are used to measure the Doppler reflex motion of the star which is induced by the presence of an orbiting planet. The amplitude of this motion is proportional to the mass of the planet $\sin{i}$, where $i$  is the orbital inclination. Therefore, only in case where we know the inclination of the orbit do we find a true mass. Typically, transit observations are used to provide this inclination.
Some RV instruments have been working for many years now, e.g. the High Accuracy Radial Velocity Planet Searcher \citep[HARPS]{Mayor03, Pepe:2002} and the northern-hemisphere counterpart, HARPS-N \citep{Cosentino2012}. Some are more recent, e.g. ESPRESSO \citep{Pepe21} and NEID \citep{2018SPIE10702E..71S}. There are also some upcoming RV instruments such as HARPS-3 \citep{2016SPIE.9908E..6FT} and KPF \citep{2016SPIE.9908E..70G}. While the technology of these instruments has improved considerably through the years, the detection of masses for very small planets still remains extremely challenging. This difficulty is primarily attributed to the effects of stellar activity on the RV signals themselves. For reference, the RV amplitude of the Earth on the Sun is 0.09\,\ms, whereas the typical stellar activity signal of the Sun is on the order of 10\,\ms, see e.g. \cite{2024MNRAS.531.4238K}. This means that with current technology we are unable to detect an Earth-analogue orbiting a Sun-like star.  

Thankfully, there has been increasing progress made on mitigating stellar activity in the extreme precision RV (EPRV) community. Some work focuses on removing the effects of stellar activity from the observed spectra  \citep{2021A&A...653A..43C,2022A&A...659A..68C}, while others focus on improving the extraction of RV information from the spectra \citep{2021MNRAS.505.1699C,2022ApJ...935...75Z,2022AJ....164...49D}. Furthermore, Gaussian Processes (GPs) are widely used to model the stellar activity in RV time-series simultaneous with the planetary signal(s) \citep[e.g.,][]{2012MNRAS.419.3147A,rajpaul_gaussian_2015,pyanetiii}. For a more complete overview of these modelling and mitigation approaches see \cite{2023AnRSA..10..623H}.

Whilst these continuing efforts of the EPRV community have enabled many more small exoplanets to gain precise mass measurements, there is a cost: each exoplanet mass is typically found using one of a variety of methods. Specifically, the choice in whether or not a planetary orbit is fixed as circular or allowed to vary in eccentricity; whether a long-term trend parameter is included; and how stellar activity in mitigated - through the used of GPs or other methods. There are also potential impacts from differing data sets with different observational sampling and cadence, and possible instrumental systematics for discussions on RV survey biases \citep{2018RNAAS...2...28M}. This inconsistency means that it is challenging to perform robust statistical analysis using exoplanet masses. By changing a few choices in the modelling of data for a single system, the extracted mass can vary significantly. 

To be able to complete statistical studies and truly understand the demographics of these systems we need a homogeneous analysis of exoplanet masses. Some recent surveys have chosen to tackle this issue as new data comes in by performing a homogeneous RV analysis, e.g. \cite{2024ApJS..272...32P}, or by designing their survey in a more unbiased way as presented by \cite{2021ApJS..256...33T}. \cite{2019ApJ...883...79D} performed a homogeneous analysis of masses (and compositions) of 11 hot-Earth planets using archival data, but since then many more small planets have been observed with radial velocities meaning this very small sample could be expanded.

In this paper we present a sample of small exoplanets where their RV observations have been analysed in a homogeneous way. This is the first time such a large scale homogeneous analysis of RV observations has been completed. We choose to focus on small planets for multiple reasons: they are most likely to be impacted by model choices/activity mitigation techniques; and they are a primary focus for upcoming missions such as Ariel, the Extremely Large Telescope (ELT) and the future Habitable Worlds Observatory (HWO).  Additionally, the internal composition of small planets is not well understood.

We focus specifically on HARPS data for several reasons: we want to have a consistent choice of instrument rather than using data from multiple sources; it is one of the top performing high resolution spectrographs which was designed for precision RV observations; and HARPS has been collecting RVs for over 20 years yielding a considerable archive of publicly available data \footnote{\url{https://archive.eso.org/scienceportal/home}}.
In an ideal world, there would be one set `best method' for modelling exoplanet RVs, however much work is still ongoing on this topic and the community as a whole has yet to reach a consensus. Instead, we present here a variety of modelling choices which are commonly used by the community as a comparison. We also provide recommendations for best practices of teams modelling their own RV data. Finally, we make available our entire workflow for this project, meaning that other teams can apply the procedures to their own data sets, or complete their own homogeneous analysis of the same data but using their method of choice. The final set of planet masses and a new, homogeneously-derived mass-radius diagram for small planets will be presented in Paper II, Osborne et al., in prep.

\section{Sample selection}
\label{sampleselection}

To reach our aim of producing a homogeneously derived sample of small planet masses we start by using the NASA Exoplanet Archive\footnote{\url{https://exoplanetarchive.ipac.caltech.edu/} Accessed 24-01-2024}. We query the archive for all confirmed planets with a radius less than 4\,\re\ and a declination below +20 degrees, taking the default parameters. We note that individual systems within the Archive often have multiple published solutions, we choose to take the default values at this stage for simplicity. We cut on declination even though this will be done implicitly when we cross-reference with the HARPS archive, however it significantly reduces the number of systems we have to cross-check. This leaves us with 1770 planets.

The next stage is to check which of these possible targets has RV data available in the HARPS public archive. There are some challenges with such a large archive spanning several principle investigators and many observing seasons, including instrument upgrades. In particular, inconsistent naming of targets makes it difficult to accurately assess how many observations each star has. To overcome this we used the catalogue of HARPS observations in \cite{barbieri2023esoharpsradialvelocitiescatalog} who were able to construct a table of HARPS RVs for the entire 20 years where they had checked the coordinates of individual systems to be able to properly match up any variation in naming.

The final sample was made by cross-referencing our targets from the NASA Archive with those of the HARPS archive. We removed any individual observations which had an SNR less than 25 and also set a minimum threshold of at least 50 HARPS observations - this is to ensure we have sufficient data epochs to perform Gaussian Process (GP) regression. For some targets there is a large amount of data available but it is from observations of transits - typically for studies of planet obliquity \citep[e.g.][]{2024A&A...690A.379K}. In these cases many observations are taken over the course of one night and so the total number of observations appears much higher, but the phase coverage of these observations is not as good.  Additionally, the Rossiter-McLaughlin effect would have to be modelled for these data \citep[]{1924Rossiter,1924McLaughlin},  which would unnecessarily increase the complexity of our models. Therefore we remove such data before modelling (see \ref{removal_rvs} for details). In the case of TIC\,301289516, the removal of the in-transit data means we are left with only 35 RV observations, which is below our minimum threshold for modelling. Therefore we remove this target from the sample.
From this, we have our final list of 87 small planets orbiting 44 stars.
The total number of planets orbiting the 44 target stars is 113, however the extra 26 are not in the small planet range (or do not have a published radius); we account for these planets in our modelling but not in our model comparison analysis.

\begin{figure}
    \centering
    \includegraphics[width=\linewidth]{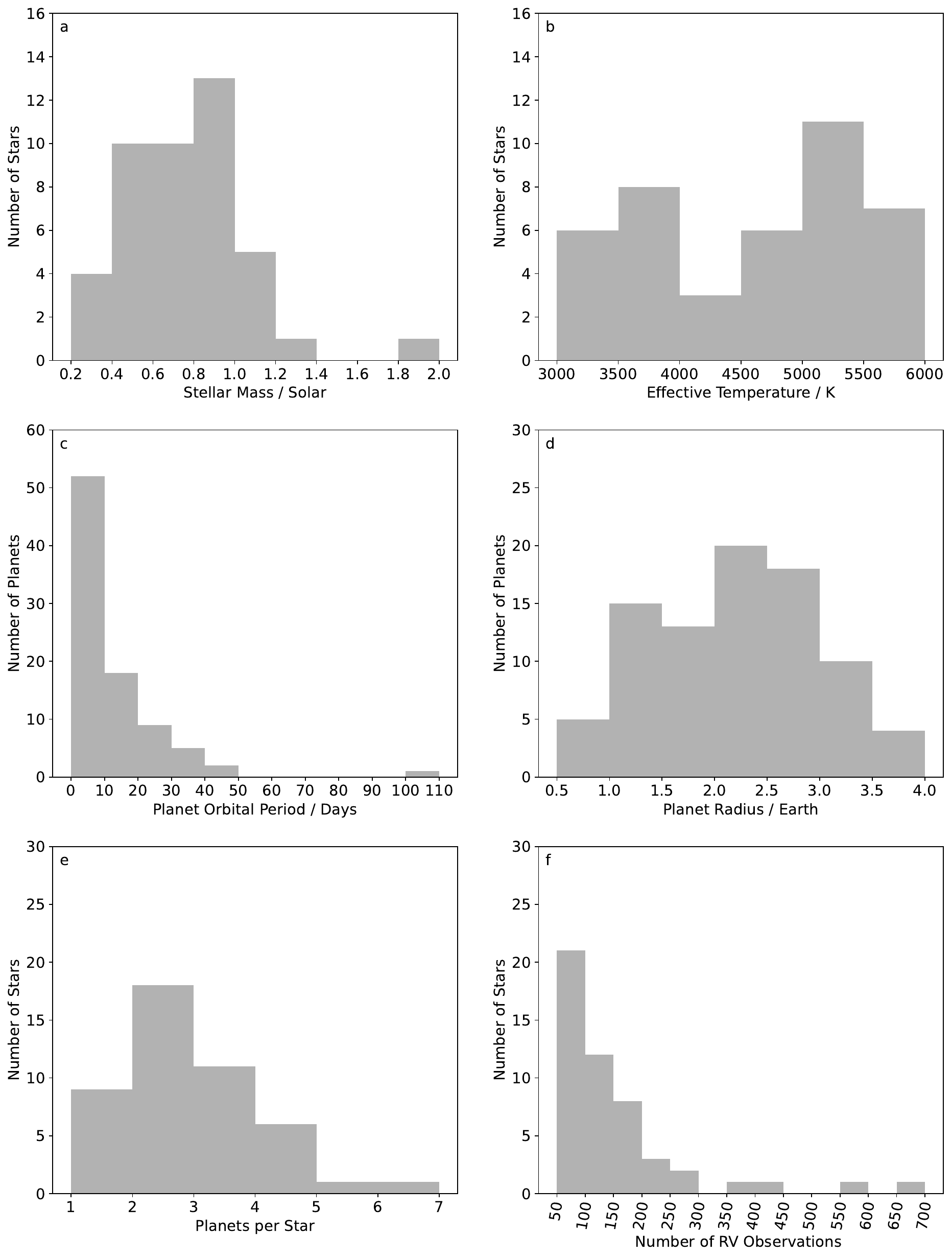}
    \caption{Histograms showing the properties of the stars and planets in our sample: a) the stellar mass, b) the stellar effective temperature, c) the planet orbital period, d) the planet radius, e) the multiplicity (number of planets per star), and f) the number of RV observations per system.}
    \label{fig:sample}
\end{figure}

Fig. \ref{fig:sample} shows histograms of the targets in our sample. Panel a) shows the distribution in effective temperature of the 44 stars, and panel b) the stellar mass. These plots show that the sample covers a fairly wide range of stellar types, with peaks around M-dwarfs and K-dwarfs. These are often specifically targeted in RV surveys as the amplitude of Doppler reflex motion of planets around less massive stars is relatively larger than for the same planet around a more massive star. The stars in the sample are all brighter than V magnitude 15, and have a median brightness of V magnitude 11. The stars are uniformly distributed across the southern sky with declination ranging from +10 to -80 degrees.
In panel c) we show the distribution of orbital periods of the planets in our sample, and in panel d) the planet radii. The majority of planets have short ($<$ 20 days) orbits, this is expected as these planets are by far the easiest to detect in transits. The distribution of planet radii peaks at around 2.0 - 2.5\,\re\ and drops off toward larger radii - as seen in demographic studies of small planets. The lack of planets at very short radii ($<$ 1\,\re) is likely due to observation biases. 
Interestingly, the radius valley is seen in our sample between 1.5 and 2.0\,\re, even with a relatively small sample. Our sample of planets is therefore a reasonable representation of the wider distribution of small exoplanets found in demographics studies, see \cite{2021ARA&A..59..291Z} for a review of exoplanet statistics.  
We also show the number of planets per target star in our sample, in panel e). The majority of planets in our sample are in multi-planet systems, most commonly in a two-planet system.

Finally, in panel f), we show the number of observations per star available within the HARPS data. The majority of our targets have below 200 observations, however 9 targets have more than this, with one target having over 650 epochs of RV observations.

\section{HARPS observations}
\label{datareduction}
For our analysis we use RV measurements and activity indicators from the HARPS spectrograph, based on publicly available reduced data from the ESO archive, as described in \cite{barbieri2023esoharpsradialvelocitiescatalog}. 

HARPS is a stabilised high-resolution spectrograph with resolving power of 110000, capable of sub-{m\,s$^{-1}$} RV precision for bright, slow rotating stars. The instrument is mounted on the ESO~3.6\,m telescope at the La Silla Observatory in Chile. The observations used in this study all use the high-accuracy mode with a 1\,arcsec science fibre on the science target. The second fibre can be used for simultaneous wavelength calibrations. The stars in this sample span a wide range of magnitudes and thus exposure times from a few minutes up to an hour.

The data available through the ESO archive has been processed using the online HARPS pipeline \citep{Pepe:2002}, and includes extracting 2D spectra that are flux correction to match the slope of the spectra across echelle orders. RV information are extracted from the spectra by cross correlating with a binary mask matching the stellar type of the star \citep{Baranne:1996}. The SNR cut at 25, enforced when cross referencing with the list of small, known planet from the NASA Exoplanet Archive, entails most RVs have precisions of 1-5\,\ms. The median RV uncertainty of this entire dataset is 1.7\,\ms, but this increases up to 12.2\,\ms for some systems. We do note that a few individual data points appear to have very large uncertainties, it is possible that their SNR is incorrectly labelled in the database and so they are not removed by the cut on SNR as they should be, however these will not significantly affect our results. In the later analysis, we use activity indicators based on the CCF bisector-span (BIS), and CCF-FWHM, as we find these are more robust for the largest sample of spectra than chromospheric activity indicators such as Ca II H\&K, and $\mathrm{H\alpha}$ which may be more sensitive to low data fidelity.

The only additional step to the data collection was to also do a sigma-clip on the RV data. We used the \texttt{astropy} sigma-clip tool to cut out any RVs more than 3 sigma away from the median, with a maximum of 2 iterations of this. This finds the median and standard deviation of the data and then removes points which are more than 3 standard deviations above our below that median value. This is done for two iterations. For most targets, the sigma-clipping does not remove any data because the standard deviation is large anyway. For a few targets with many (>500) observations, it does remove more observations, but still leaves most for the modelling. For one single target, we have only 49 RV observations left after sigma-clipping, this is below our original threshold of 50 epochs but we choose to include it in the sample anyway.

For a few targets we did also need to remove specific data points which were clear outliers, see \ref{removal_rvs} for details of this. After the sigma clipping and removal of data for specific targets we are left with 6428 individual RV observations. The baseline of observations ranges from approximately 60 to 6700 days, with a median baseline of around 600 days. The full data used for the final fits for all targets is available online.

It is becoming increasingly common in RV analysis to use Gaussian processes (GPs) to help mitigate against stellar activity \citep[e.g.,][]{2014MNRAS.443.2517H,2015ApJ...808..127G}. One promising way to do this is a multi-dimensional GP where you fit both the RV data and some activity indicators simultaneously\citep[for an explanation of this process see, e.g.,][]{pyanetii,2023ARA&A..61..329A}. There are a range of different activity indicators which can be used. In our analysis we chose to include some of the most commonly used ones: the full width at half-maximum (FWHM), the bisector inverse slope (BIS).

While we have chosen to use HARPS data to ensure the consistency in the RV instrument used, given the long time HARPS has been observing there are some offsets in the data. In particular, the HARPS fibre was upgraded in 2015, causing a possible RV offset. It is recommended to account for this shift by modelling the HARPS pre-upgrade data as from a different instrument to the HARPS post-upgrade data. For this reason, we label all RV data as either `HARPS$\_$pre' or `HARPS$\_$post'. Within the modelling we set these as two separate instruments - meaning we account for an offset between them. In most cases the data is all after the upgrade, but in a few cases the data is from before the upgrade, or contains a mixture of pre- and post- fibre upgrade.

\section{Automating the process}
\label{automating}

One of the biggest challenges of this project was in the initial setup of the RV modelling. We wanted to move away from an `artisanal' approach of looking at one system at a time, to automating a process to model many systems at once.  This is for two reasons: we want our method to be generally applicable to any system; and we did not want to introduce any biases by manually setting the priors and input parameters.

To overcome this we first query the NASA archive to find the required parameters for each individual target: the stellar mass, radius, and temperature, and the planet period, transit midpoint, radius, and mass. We use the composite parameters table from the archive for these values. The planet orbital period and transit midpoint (for non-transiting planets this parameter has the same name but refers to the time of conjunction) are used as priors in our RV modelling. The other parameters are used for comparison in our results and discussion, but not used directly in the modelling. 
We separate the HARPS archive data into individual files for each star, applying our sigma-clipping and removal of specific data where necessary. For each model we create an input file for the fitting toolkit which can be used for all targets i.e. we have 1 input file per model not per target. This significantly reduces the time needed to set up each model run. 

\section{RV modelling choices}
\label{modellingchoices}

Distinguishing between planetary and stellar signals in RV data remains a challenge in exoplanet detection. The task is particularly difficult when dealing with active stars, where stellar activity can produce RV signals that mimic or obscure those of orbiting planets. Gaussian Processes (GPs) have emerged as one of the most powerful tools to address this issue. GPs offer a highly flexible and non-parametric approach to modelling complex stochastic variations, such as those induced by quasi-periodic stellar activity using tailored quasi-periodic kernels \citep[see][]{2023ARA&A..61..329A}.

The benefits of using GPs extend beyond their flexibility. They can incorporate prior knowledge about the system, such as the expected periodicity of stellar activity, and can be combined with other models to account for additional physical processes. This adaptability makes GPs particularly well-suited for analysing spectroscopic time series of active stars, where traditional methods often struggle. Indeed, numerous studies have demonstrated the effectiveness of GPs and their variants, such as multidimensional GPs, in identifying RV signals of planets in the presence of significant stellar activity \citep[see][]{Rajpaul2015,pyanetiii}.
The multidimensional variant of GPs has demonstrated its effectiveness in enhancing the precision of planetary detection, particularly in scenarios where activity indicators provide significant information about the underlying stellar signals \citep[see][]{Barragan2023}. This approach leverages the underlying relation between RV data and activity indicators, allowing for a more accurate separation of planetary and stellar signals.
However, the advantages of this multidimensional GP framework diminish under certain conditions. Specifically, when the data suffers from sub-optimal sampling or is dominated by large amounts of white noise, the activity indicators may fail to capture information about the stellar signals. In such cases, the use of a multidimensional GP does not offer any significant improvement over traditional methods, as the lack of reliable activity information limits the framework's ability to accurately model the stellar signal \citep[see][]{Barragan2024}.

A commonly employed kernel that allows modelling stochastic periodic behaviour is the Quasi-Periodic (QP) kernel \citep[as introduced by][]{Roberts2013}, defined as
\begin{equation} 
\gamma_{\rm QP}(t_i,t_j) = A^2 \exp \left\{ - \frac{\sin^2\left[\pi \left(t_i - t_j \right)/P_{\rm GP}\right]}{2 \lambda_{\rm p}^2} - \frac{\left(t_i - t_j\right)^2}{2\lambda_{\rm e}^2} \right\} \label{eq:qpkernel}, 
\end{equation}
\noindent where $A$, the amplitude, is a parameter that works as a scale factor that determines the typical deviation from the mean function, $P_{\rm GP}$ represents the characteristic periodicity of the Gaussian Process, $\lambda_{\rm p}$ denotes the inverse of the harmonic complexity (indicating the complexity within each period), and $\lambda_{\rm e}$ represents the timescale of long-term evolution.

Once we have our data files for each system we then have to choose how we model the RVs to find the planet masses.
We chose to use the package $\texttt{Pyaneti}$\citep{pyanetii,pyanetiii} for all of our modelling as this has a variety of options available for the fitting and is partly written in fortran90 meaning it runs much faster than an entirely python-based code. 
$\texttt{Pyaneti}$is also a fairly common choice of package within the RV modelling community and makes use of multi-dimensional GPs \citep{rajpaul_gaussian_2015,Jones2017,Gilbertson2020} to mitigate stellar activity. For a full description on how $\texttt{pyaneti}$ implements the QP kernel described in \eqref{eq:qpkernel} within the multi-GP framework see \citet{pyanetiii}.
Other RV fitting toolkits which make use of multi-dimensional GPs include PyORBIT \citep{pyorbit1,pyorbit2} and S+LEAF \citep{Sleaf}. 
In addition to our goal of providing a homogeneously derived sample of small planet masses, we also wanted to investigate how the choices in modelling affect the derived planet mass. 
For this reason we chose to run twelve different models on the data. We wanted to compare the impact of: using a GP versus no GP; the dimension of GP used; adding a long-term trend parameter; and modelling orbits as circular or eccentric. See Table \ref{tab:models} which outlines all the models we used.

For all runs, we perform Markov chain Monte Carlo (MCMC) samplings using the sampler included in \texttt{pyaneti} based on ensemble sampler \citep{foreman-mackey_fast_2017}. We sample the parameter space with 200 Markov chains. Each chain is initiated randomly with values within the prior ranges.  
We create posterior distributions with the last 1000 iterations of converged chains with a thin factor of 10. This generates distributions with 200000 independent points per each sampled parameter.

\begin{table*}[]
\centering
\caption{Overview of the 12 models applied to RVs and activity indicators in this study.}
\label{tab:models}
\begin{tabular}{llllll}
\hline
Model & GP Dimension & Orbital Eccentricity         & Long-term trend        & Activity Indicator & GP Kernel \\ 
\hline
a & 0            & circular            & no trend           & none               & none                             \\
b & 0            & circular            & linear trend       & none               & none                             \\
c & 0            & circular            & quadratic trend    & none               & none                             \\
d & 0            & eccentric           & no trend           & none               & none                             \\
e & 0            & beta distribution   & no trend           & none               & none                             \\
f & 1            & circular            & no trend           & none               & Quasi-Periodic                   \\
g & 1            & eccentric           & no trend           & none               & Quasi-Periodic                   \\
h & 2            & circular            & no trend           & FWHM               & Quasi-Periodic                   \\
j & 2            & eccentric           & no trend           & FWHM               & Quasi-Periodic                   \\
k & 3            & circular            & no trend           & FWHM+BIS           & Quasi-Periodic                   \\
m & 3            & eccentric           & no trend           & FWHM+BIS           & Quasi-Periodic                   \\
n & 3            & beta distribution   & no trend           & FWHM+BIS           & Quasi-Periodic                   \\

\hline
\end{tabular}
\\
\footnotesize{Each row represents a single model, with the columns indicating the choices made for that model. Note: we do not label models as i or l to avoid confusion with the number 1.}
\end{table*}

As well as the specific model choices, we also wanted to be consistent in our application of priors for the modelling parameters. We chose to set a Gaussian prior on both the orbital period, $P$, and time of conjunction, $T_c$, listed in the NASA archive, using the 1$\sigma$ errors. Typically these values have been found through transit fitting of the planets. For several systems there are no listed values of this on the archive, for these we manually check the original publications and add in the values ourselves - details are in Appendix \ref{manualpriors}. For all other planetary orbit parameters we chose to use wide uniform (un-informative) priors. For the eccentricity of the planetary orbit we set either a fixed at zero eccentricity (for our circular model cases), or parameterise the eccentricity and argument of periastron to
\begin{equation}
\label{equation}
    e\omega_1 = \sqrt{e\sin{\omega_*}} \text{  and  } e\omega_2 = \sqrt{e\cos{\omega_*}}.
\end{equation}
This has the benefit of not truncating at zero which is often a problem in modelling eccentricities \citep{1971Sweeney}. However, after running models including eccentricity we noticed that in some cases there is a very high eccentricity found which seems unlikely for so many systems. This is likely due to the model fitting high eccentricity orbits to spurious outliers in the data \citep{Hara2019}. For this reason we also chose to run two additional models (models e and n, as described in Table \ref{tab:models}) which puts a prior on the eccentricity as a beta distribution. We use the form of \cite{VanEylen19} for single-planet systems, as this is the more general case.

For the GP hyperparameters we again use wide uniform priors. Except in the case of the GP period, $P_{GP}$, where we set this based on the stellar type.
It has been shown that the GP period links strongly to the stellar rotation period \citep{Nicholson2022}. For each star in our sample, we use the published stellar effective temperature and convert this to a B-V magnitude, then using the relation from \cite{2008ApJ...687.1264M} we can estimate the maximum stellar rotation period for a given stellar age. Taking the upper limit of 9 Gyr, we assign maximum rotation periods of 60, 50 , 40, and 20 days for stars with temperatures < 4000K, 4000 - 5000K, 5000 - 6000K, and > 6000K, respectively. We also then set the maximum timescale of evolution of active regions, $\lambda_\mathrm{e}$, to be twice this rotation period. We note that future work may benefit from using more physically-motivated GP hyperparameter priors based on stellar type and age.

The choice of priors for the multi-GP amplitudes was informed by the results of previous analyses that reflect the underlying correlations between the RVs and the CCF-derived activity indicators. \citep[e.g.,][]{Barragan2019,pyanetiii,Barragan2023}. Specifically, previous studies observe that when the RV amplitudes ($A_0$ and $B_0$)are positive, the corresponding amplitudes for the FWHM are also positive, while those for the BIS are negative. For this reason we set $A_0$ and $B_0$ to be positive, and we leave the amplitudes for the other hyper-parameters to be vary more.

\begin{table*}[]
\caption{Summary of priors used for all parameters in all models.}
\label{tab:priors}
\begin{tabular}{p{0.3\linewidth} p{0.2\linewidth} p{0.4\linewidth}}
\hline
Parameter & Prior & Notes \\
\hline
Orbital Parameters & & \\
\hline 
Mid-transit time, T$_0$, days & $\mathcal{N}[a,b]$ & Where $a$ is the mid-transit time from the archive, and $b$ is the uncertainty on that time. \\
Period, P, days & $\mathcal{N}[c,d]$ & Where $c$ is the period from the archive, and $d$ is the uncertainty on that period. \\
eccentricity, e & $\mathcal{F}[0]$ & For the circular model. \\
argument of periastron, omega & $\mathcal{F}[0]$ & For the circular model. \\
ew1 & $\mathcal{U}[-1,1]$ & For the eccentric model. \\
ew2 & $\mathcal{U}[-1,1]$ & For the eccentric model. \\
RV amplitude, k, km/s & $\mathcal{U}[0,0.5]$ & \\
\hline 
GP Hyperparameters & & \\
\hline
A0, \kms & $\mathcal{U}[0,0.5]$ &  \\
B0, \kmsd & $\mathcal{U}[0,0.5]$ &  \\
A1, \kms & $\mathcal{U}[-0.5,0.5]$ &  \\
B1, \kmsd & $\mathcal{U}[-0.5,0.5]$ &  \\
A2, \kms & $\mathcal{U}[-0.5,0.5]$ &  \\
B2, \kmsd & $\mathcal{U}[-0.5,0.5]$ &  \\
Timescale of active regions, $\lambda_\mathrm{e}$, days & $\mathcal{U}[1,160]$ & The upper limit is set to twice the period of the GP\\
Inverse of Harmonic Complexity, $\lambda_\mathrm{p}$ & $\mathcal{U}[0.01,2]$ &  \\
Period of GP, P$_{GP}$, days & $\mathcal{U}[0,80]$ &  This is set based on the stellar effective temperature \\
\hline
\end{tabular}
\end{table*}

\section{Results and discussion}
\label{results}

We have re-modelled 6428 HARPS RV measurements for 44 stars harbouring 87 small, transiting planets. In this section, we summarise our findings and analyse the impact of model choice when fitting RV signals. For three of our targets, TOI-269, TOI-4399, and HD\,3167 our models could not provide a good fit to the data available. TOI-269 is an active M dwarf star where a custom RV extraction was used in the discovery paper alongside additional photometric data to provide a good fit \citep{2021Cointepas}. TOI-4399 is a very young star with strong activity signals and no published mass measurement \citep[only an upper limit,][]{2022Zhou}, our modelling suggests that this system requires additional data to characterise the planetary mass fully. HD\,3167 has only 50 RV observations but is a 4-planet system; leaving only a low degree of freedom when fitting 20+ parameters, depending on the model-choice. When modelling this system with a GP, this issue is amplified and the degrees of freedom is too low to fit the data well. For the following sections, we remove these three target stars from our analysis, resulting in a total of 83 small planets orbiting 41 stars. For completeness, the final results tables do include the fits for these three planets.

The extracted RV amplitude, eccentricity, BIC, and AIC for all models for each planet are shown online at \href{https://doi.org/10.5281/zenodo.14170646}{Table B.2}. The full posterior distributions and plots of each system, plus the full list of best-fit parameters and GP hyperparameters are available online. The masses given in the `params' file in these full results online should be interpreted as an $m\sin{i}$ value, and we note that the stellar parameters used to calculate these are from the default NASA Exoplanet Archive table, meaning they are heterogeneous in nature themselves. Paper II will provide a more homogeneous set of planet masses using a consistent stellar characterisation method. Here we summarise the main findings by comparing the impact of different modelling choices.

\subsection{Impact of long-term trends}

\begin{figure*}
    \centering
    \includegraphics[width=0.8\linewidth]{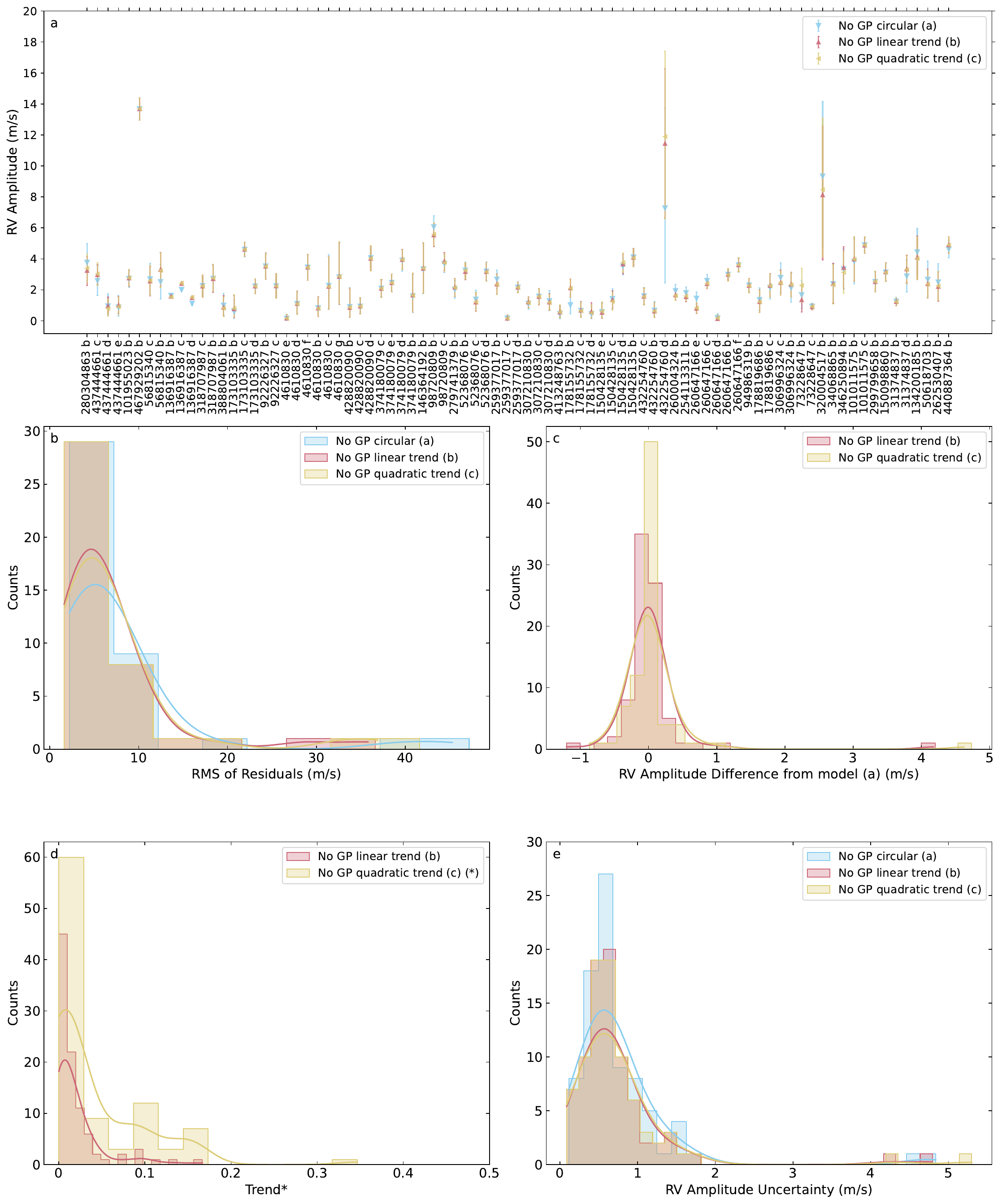}
    \caption{Impact of long-term trends. Panel a) Comparison of the RV amplitude found for each target for three different models, (a), (b) and (c), error bars show the 1-sigma uncertainty from the MCMC posteriors. Target names are given as TIC IDs with the letter of the planet. Panel b) histogram showing the root-mean-squared error of the residuals to the fit. Panel c) histogram showing the RV amplitude found compared to model (a). Panel d) histogram showing the amplitude of the trend found (for the linear trend in \ms\,days$^{-1}$, for the quadratic trend in \ms\,days$^{-2}$), * the value for the quadratic trend amplitude has been multiplied by 1000 to plot on the same axes. Panel e) histogram showing the 1-$\sigma$ uncertainty in the RV amplitude found for the different models.}
    \label{fig:trends}
\end{figure*}

We compare the extracted $K$ amplitude (RV amplitude) for each target with the different models. Figure \ref{fig:trends}, panel a) shows the extracted $K$ amplitude for the 3 models (a, b, c) which have no GP added to mitigate stellar activity. The difference between the models is that b has a long-term linear trend added, c has a long-term quadratic trend added, and a has no long-term trend. The purpose of adding a long-term trend in RV modelling is typically to account for potential changes in the instrument/telescope over long baselines or to account for the impact of a longer-period unknown planet (or star) in the system \citep[e.g.][]{2019Espinoza,2023Korth}. In some cases, a significant measurement of a long-term trend in RV data has been used to claim the discovery of a planet candidate \citep{2022Lubin}. We wanted to test whether adding a model of a long-term trend to all systems, regardless of whether we think there is a potential for an additional planet, makes a difference to the extracted RV amplitude. In Fig.\,\ref{fig:trends}, panel a) we show, for most targets, the addition of any long-term trend makes only a very small difference to the $K$ amplitude found. This is likely because no trends are evident in the data for these targets. However, in a few cases, a more noticeable difference is seen, and although the error bars typically overlap, the median amplitude found can vary by 1\,\ms\ or more. The difference between a linear and quadratic long-term trend is very minor, the 1$\sigma$ error bars overlap almost completely for all planets. 

In Fig.\,\ref{fig:trends}, panel b) we show a histogram of the root mean squared (RMS) of the residuals for the three models (a, b, and c). The highest RMS of residuals is for model a, with no long-term trend added. The overall distribution is very similar for all 3 models. 

Fig.\,\ref{fig:trends}, panel c) shows a histogram of the difference in RV amplitude of our models compared to the most simple model i.e. the RV amplitude for model c) subtract the RV amplitude for the same planet for model a). In both the linear and quadratic trend cases, the distribution centres around 0\,\ms, and almost all targets have a difference within $\pm$\,1\,\ms.

The amplitude of the trend itself for both the linear and quadratic cases is shown in Fig.\,\ref{fig:trends}, panel d). Note that the quadratic trend case has been multiplied by 1000 to allow it to be visible on the same axes. 
For the linear case first (model b), the amplitude of the trend is below 0.20 \ms\,days$^{-1}$ in all cases with most targets having lower than 0.05 \ms\,days$^{-1}$. For the quadratic trend case, all targets have trend amplitudes below 0.35 $\times 10^{-3}$ \ms\,days$^{-2}$, with almost all targets less than half this amount. 
Based on these results, the indiscriminate addition of a long-term trend to the model does not make a significant difference on the planet mass found because the amplitude of the trend is very small. 

Finally, in Fig.\,\ref{fig:trends}, panel e) we show the distribution of RV amplitude uncertainties for the 3 models (a, b, and c). the peak of the distribution in all cases is around 0.6\,\ms, with the highest uncertainties of around 5\,\ms. The difference between the 3 models is not significant, it is likely that the systems in our sample do not have an external high-mass companion, or the RV observations we have are not sensitive to the long-term trend.

\subsection{Impact of eccentricity}

\begin{figure*}
    \centering
    \includegraphics[width=0.8\linewidth]{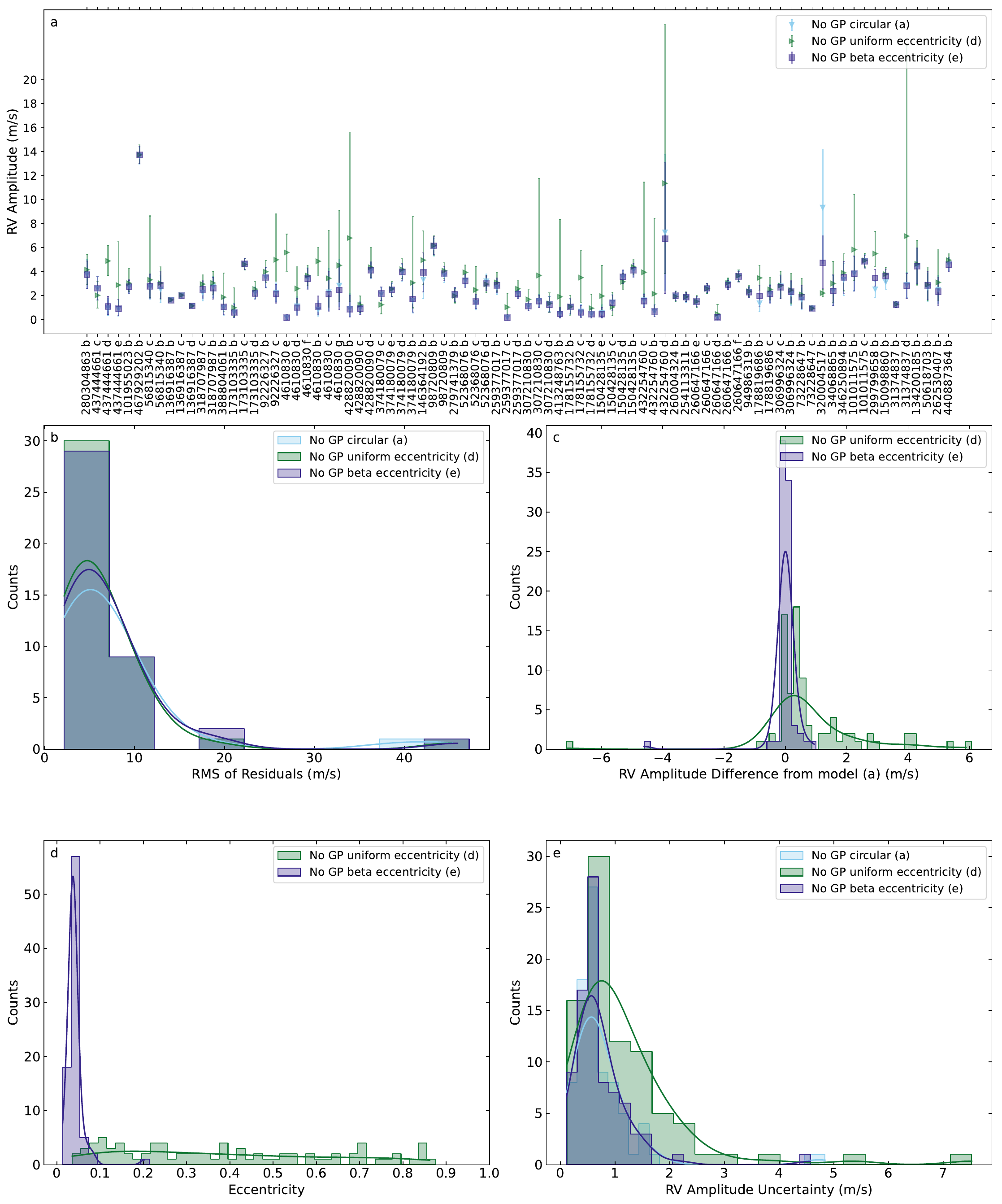}
    \caption{Impact of orbital eccentricity. Panel a) Comparison of the RV amplitude found for each target for three different models, (a), (d) and (e), error bars show the 1-sigma uncertainty from the MCMC posteriors. Target names are given as TIC IDs with the letter of the planet. Panel b) histogram showing the root-mean-squared error of the residuals to the fit. Panel c) histogram showing the RV amplitude found compared to model (a). Panel d) histogram showing the value of eccentricity found. Panel e) histogram showing the 1-$\sigma$ uncertainty in the RV amplitude found for the different models.}
    \label{fig:noGP}
\end{figure*}

When investigating how modelling orbits as circular or eccentric impacts the planet masses found, we take the no-GP model and either set a uniform eccentricity prior, model d, or we set the prior on eccentricity in the form of a beta distribution, model e. Both of these cases are described in Section \ref{modellingchoices}. Fig.\,\ref{fig:noGP} shows the results of these models compared to the circular case (model a). 

In Fig.\,\ref{fig:noGP}, panel a) we show how the RV amplitude changes for the 3 models for each planet in our sample. For some very small planets, the eccentric case gives an RV amplitude which is significantly different from the circular or beta distribution case, e.g. for TIC\,437444661\,d, and TIC\,4610830\,f. Even for planets where the 1-$\sigma$ error bars overlap, the different in the median value of RV amplitude varies by as much as a factor of 3, e.g. for TIC\,428829990\,d it ranges from approximately 2\,\ms\ to 7\,\ms. For all planets, models a), circular, and e) beta distribution, give the most similar results, with model d) eccentric, giving the most different ones. Panel b) in Fig.\,\ref{fig:noGP} shows the RMS of the residuals for the 3 models. The distributions for all 3 models are very similar with no significant differences between them.

When comparing the RV amplitude found for models d) and e) compared to model a), showed in panel c) of Fig.\,\ref{fig:noGP}, we find that a uniform prior on the eccentricity (model d) gives, on average, higher values of RV amplitude - and therefore higher planet masses. For this eccentric case, the RV amplitude difference centres slightly offset from 0\,\ms\ and has a much wider range, up to around $\pm$ 6\,\ms. For the beta distribution case, model e), the amplitude difference centres around 0\,\ms\ and has a much smaller range of values, with almost all planets having less than 1\,\ms\ difference in RV amplitude compared to the circular case. The model with a beta distribution on eccentricity gives much more similar RV amplitudes to the circular case, whereas the uniform prior on eccentricity gives higher RV amplitudes on average. This highlights the importance on choosing the prior on eccentricity with care. 

We also perform this analysis for the 3D GP models which have different eccentricities - the circular case (model k), uniform prior on eccentricity case (model m), and the beta distribution on eccentricity case (model n). We find a very similar results - the beta distribution gives the most similar results to the circular case. The model with a uniform prior on eccentricity tends to find higher RV amplitudes on average. 

The distributions of eccentricity values found for the models with a uniform prior on $e$ (model d) and a beta distribution (model e) are shown in Fig.\,\ref{fig:noGP}, panel d). For model d), the distribution is almost flat, with eccentricity values ranging all the way up to 0.9. For the beta distribution case, model e), the eccentricities found centre close to zero, with the highest value being 0.2, following fairly closely the prior distribution set. Given the very high values of eccentricity found in the case for model d), we wanted to check that the wide priors set on the RV amplitude were not contributing to this. We ran model d) again for all targets but restricted the RV amplitude to be less than 50\,\ms. We found that the eccentricity and RV amplitude did not change by more than 1$\%$ in any case and so the wide priors on RV amplitude are not the reason for the high eccentricities. In some ways it is surprising to find such high values of eccentricity, especially as we chose to parametrise the eccentricity and argument of periastron as in eq. \ref{equation}, which should help with this issue. It is possible that the model is finding such high eccentricities due to spurious data points \citep{Hara2019}.

Finally, Fig.\,\ref{fig:noGP}, panel e) shows the RV amplitude uncertainty for each planet found with models a), d) and e). The circular and beta distribution models, a) and e), have the most similar RV uncertainties. Both with distributions peaking below 1\,\ms\ and only a few higher outliers. For the eccentric case, model d), the RV distribution in RV amplitude uncertainty peaks at a higher value and has higher outliers - up to 7\,\ms.

Based on our results, it is clear that using a uniform prior on eccentricity is not a suitable approach for modelling large sets of RV data. Instead, we suggest using an informative prior distribution on the eccentricity, such as a beta distribution. We note that the RV amplitude (and therefore planet mass) found for the whole sample does not change much between just fixing the orbits as circular compared to the beta distribution in eccentricity. However we believe that the use of beta distribution is more physically motivated as we would not expect every planet in our sample to be on a perfectly circular orbit. Alternatively, the simultaneous modelling of photometric data may help accurately constrain the eccentricity, though testing this in more detail is beyond the scope of this paper.

\subsection{Impact of GP dimension}

We compare the $K$ amplitude found with different dimensionalities of GP - with no GP, model a; a 1-dimensional GP (fitting just to the RVs), model f; a 2-dimensional GP (fitting to the RVs and an activity indicator, in this case the FWHM), model h; and  a 3-dimensional GP (fitting to the RVs and two activity indicators, in this case FWHM and BIS), model k. In all cases the models are for a circular orbit.

\begin{figure*}
    \centering
    \includegraphics[width=0.8\linewidth]{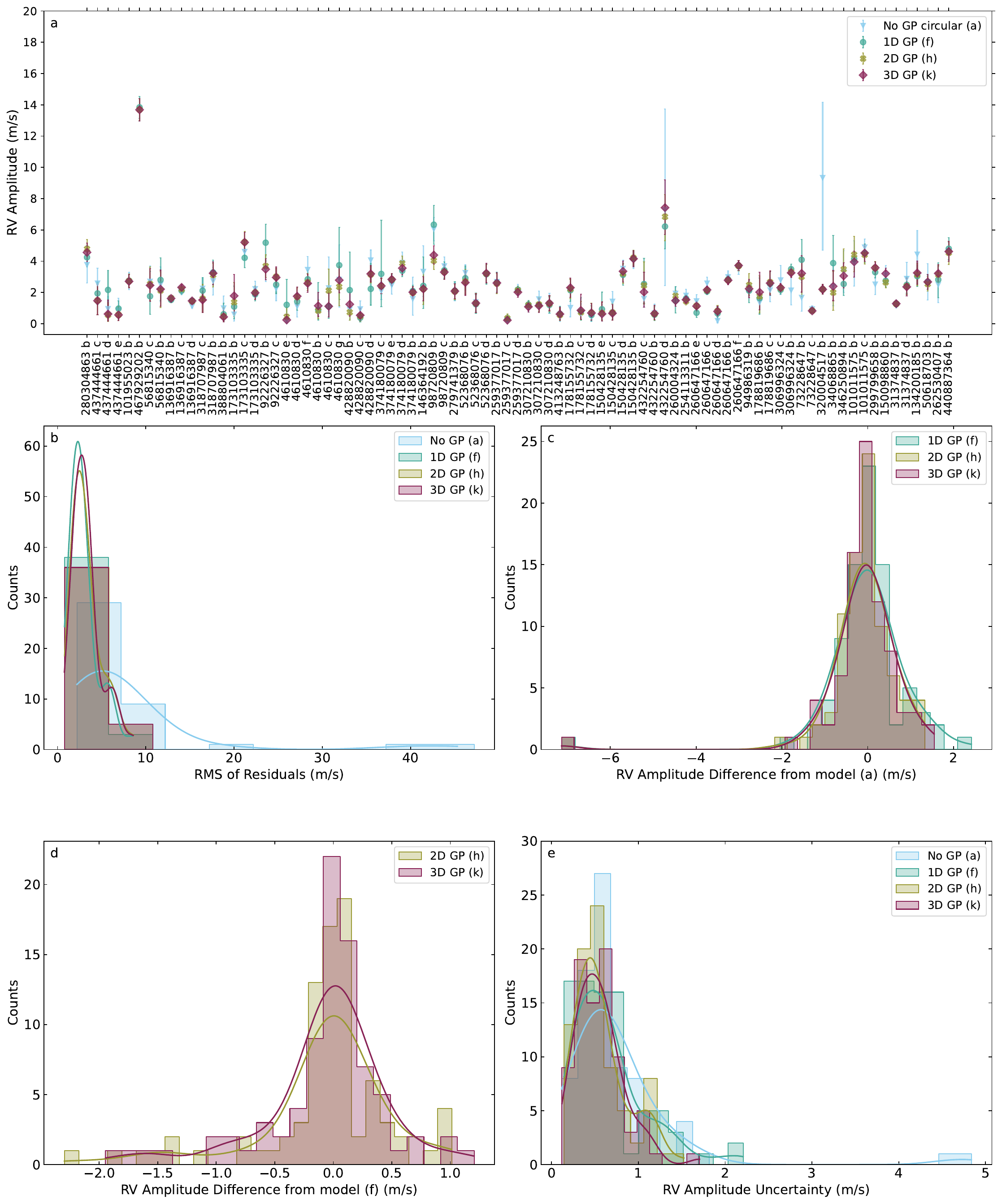}
    \caption{Impact of GP dimension. Panel a) Comparison of the RV amplitude found for each target for four different models, (a), (f), (h) and (k), error bars show the 1-sigma uncertainty from the MCMC posteriors. Target names are given as TIC IDs with the letter of the planet. Panel b) histogram showing the root-mean-squared error of the residuals to the fit. Panel c) histogram showing the RV amplitude found compared to model (a). Panel d) histogram showing the RV amplitude found for models (h) and (k) compared to model (f). Panel e) histogram showing the 1-$\sigma$ uncertainty in the RV amplitude found for the different models.}
    \label{fig:ndimensions}
\end{figure*}

Fig.\,\ref{fig:ndimensions} shows the results of these 4 models with different dimensions of GP. In panel a we compare the extracted RV amplitude for all planets in our sample with the different models. There are two things of note: the biggest error bars tend to be from the no GP case, and the biggest differences also tend to be for the no GP case. However, for nearly all the planets in the sample, the dimension of the GP does not significantly change the extracted RV amplitude. Although we do note that the median value of RV amplitude for a given planet does vary a little between the models, which would have an impact on statistical studies of the population.

In Fig.\,\ref{fig:ndimensions}, panel b we show the RMS of residuals for the 4 models. All GP models have very similar distributions, with the no GP case having the largest RMS values. Therefore, the inclusion of a GP does reduce the RMS of the residuals on average.

Fig.\,\ref{fig:ndimensions}, panel c shows the difference in RV amplitude found for each model compared to the no GP case. Here there is a slight shift seen for the 1D GP case compared to the 2D and 3D cases. The 1D GP case find slightly higher RV amplitudes on average, and is most different from the other GP models.

In fig.\,\ref{fig:ndimensions}, panel d we show the difference in RV amplitude found for the 2D and 3D GP models compared to the 1D GP case. The 2D and 3D GP models overlap very well in terms of RV amplitude. They both show some differences from the 1D GP model (which would be at 0 in this plot). The 1D GP model is the most inconsistent compared to the other two.

Finally, in Fig.\,\ref{fig:ndimensions}, panel e we show the uncertainty in the RV amplitude found for every planet with each model. All models have a peak in uncertainty at below 1\,\ms. The no GP case has a slighted shifted peak uncertainty and also has the highest outliers. The models including a GP have very similar distribution in RV amplitude uncertainty.

Based on these results we would recommend that, if using a GP, to use a multi-dimensional GP which fits to an activity indicator. This is because the 2D and 3D GP results seem the most robust compared to the 1D case; the 1D GP model find the biggest difference in RV amplitude.

\subsection{Model comparison}

We perform model comparison for each of our models by computing the Bayesian Information Criterion (BIC) and the Akaike Information Criterion (AIC). \href{https://doi.org/10.5281/zenodo.14170646}{Table B.2} gives the value of each of these for each model of every system. However, we note that none of these metrics are perfect indicators of goodness of fit, and additionally, the use of different data sets (in the case of the 2D and 3D GP models) means that you cannot directly compare these metrics. Regardless, we provide this information as an overview. For the models which use only the RV data, the lowest AIC model for 78 planets is the 1D GP circular model (f), followed by the 1D GP model with a uniform prior on eccentricity (g) for 11 planets. The remaining planets all prefer a no GP model. For the 2D GP models, 85 planets prefer the circular model (h) over the uniform prior eccentric (j). For the 3D GP case, 80 planets prefer the circular case (k), 18 the eccentric model with beta distribution (n), and 15 the uniform eccentric model (n). In general, the circular models seem to be preferred by the AIC, possibly because these have fewer parameters and so are not penalised by the goodness of fit metrics, though in some cases the non-circular models are still preferred. 

To define our `best' model for the adopted set of planet parameters for each target we first find the lowest AIC model from the models which use only the RV data (models a to f). If the best model for a given target corresponds to the 1D GP model we take this as an indication that for that specific target a GP is required. For these targets which require a GP, we assign the 3D GP with beta distribution as the best model. For the targets which prefer a no GP model, we assign the no GP beta distribution model as the best model. 

We note that the beta distribution models do not always give the `best' fit in terms of AIC and BIC. However, we choose these as our final models as the treatment of eccentricity is the most realistic: not all planets in our sample will be on circular orbits, and using a uniform prior for eccentricity gives spuriously large eccentricity values. Additionally, the beta distribution is an empiric result based on transit observations of small planets and so has a good physical motivation \citep{VanEylen19}. We choose the 3D GP case rather than the 1D or 2D case because the 1D GP case seems the least consistent with the others (in terms of extracted RV amplitude), and because the 3D GP case makes use of the most information - in the form of the FWHM and BIS indicators. We note that the 3D GP model will always have a lower value of AIC compared to the 2D GP case because it has more data points, but that is not why we choose this model. 

\begin{figure*}
    \centering
    \includegraphics[width=0.8\linewidth]{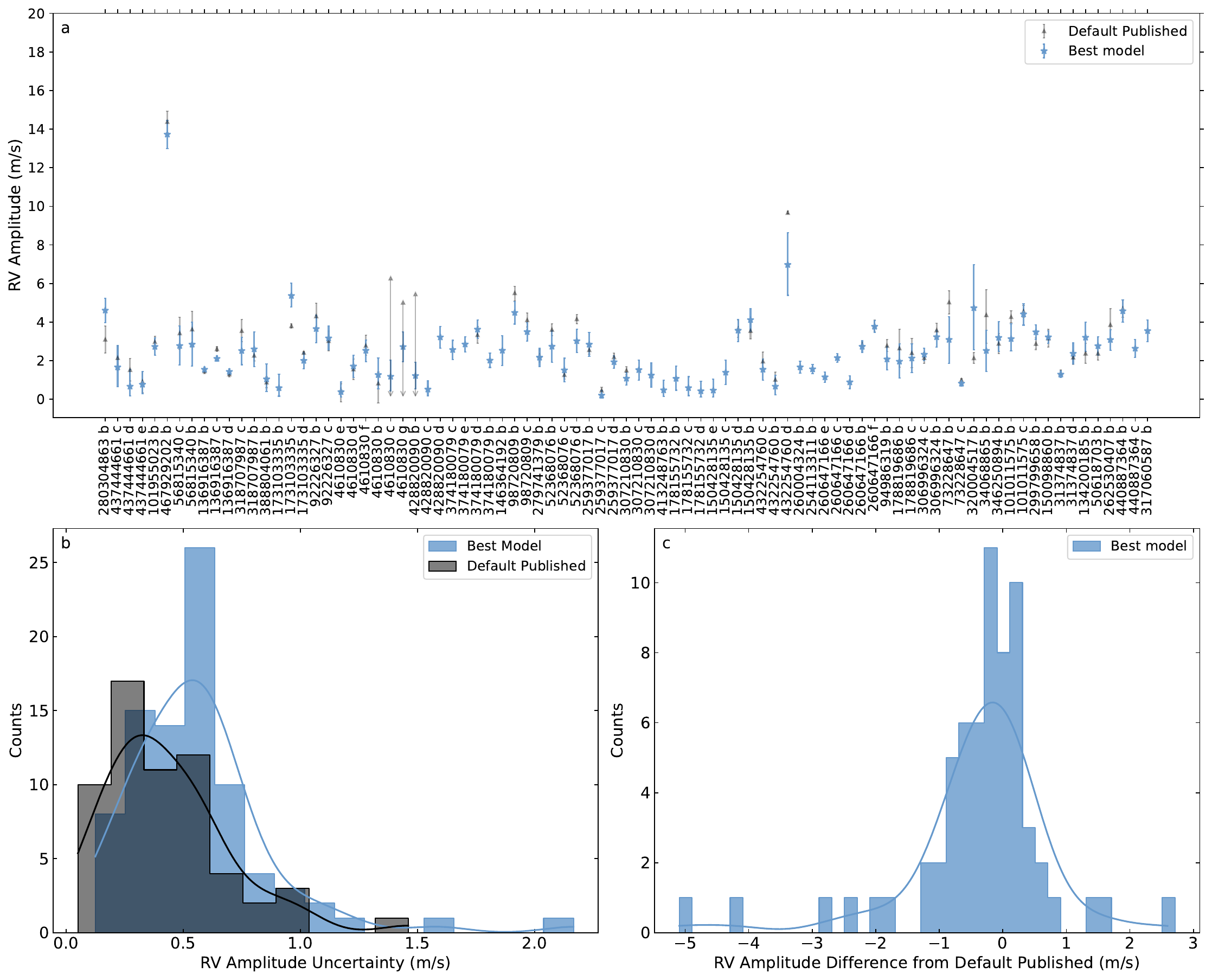}
    \caption{Panel a: The RV amplitude found with the best model for each planet in our sample, blue stars, compared to the default published value from the NASA Exoplanet Archive, grey squares. Some planets do not have a published RV amplitude on the archive. Panel b: a histogram showing the RV amplitude of our best model for each planet subtracted by the default published value. Panel c: histogram showing the RV amplitude uncertainty for our best model compared to the default published.}
    \label{fig:best_model}
\end{figure*}

Fig.\,\ref{fig:best_model}, panel a shows the RV amplitude of the best model for each planet in our sample compared to the default published value from the NASA Exoplanet Archive (though we note that for some planets they do not have a published RV amplitude and many which do have a value use different data from our work). For some planets there is a significant difference between the two. Even where the 1-$\sigma$ error bars overlap, the difference in the median RV amplitude for a given planet can vary by up to a factor of 2. This would have a big impact on the calculated bulk density of the planet and therefore change the estimated composition.

In panel b we show a histogram of the RV amplitude differences from our best model compared to the default published values (for targets which have this published value). The distribution here does peak around 0\,\ms, however some targets have difference in amplitude of up to 5\,\ms. On the one hand this is reassuring as it seems that our results are broadly consistent with the literature. On the other hand, there are still differences seen for some targets, which would have an impact on the overall demographics of this sample. This highlights the need for a homogeneous analysis approach if we want to study populations of planets rather than individual systems.

Finally, in panel c we present a histogram of the RV amplitude uncertainties for our best model compared to the default published. It is clear that the default published amplitudes have a lower uncertainty on average compared to our best model. We discuss why this may be the case in Sect.\,\ref{caveats} (most likely due to additional data being used in the published works) and note that the aim of this work was not to find the most precise planet masses, but rather to provide a sample where the masses have been found homogeneously. Overall, this comparison shows that a homogeneous approach finds a different distribution in RV amplitudes for some targets (and therefore planet mass) compared to a heterogeneous sample.

\subsection{TIC 98720809: a representative example}
In this section we show the full results for one example, TIC 9870809, a two-planet system which has a very consistent RV amplitude found across all models with a GP. For this target we take the 120 HARPS RV observations (all post fibre upgrade) and the priors found from the NASA Exoplanet Archive composite parameters table. This gives the priors listed in Table \ref{tab:priors_987}.

\begin{figure*}
    \centering
    \includegraphics[width=0.7\linewidth]{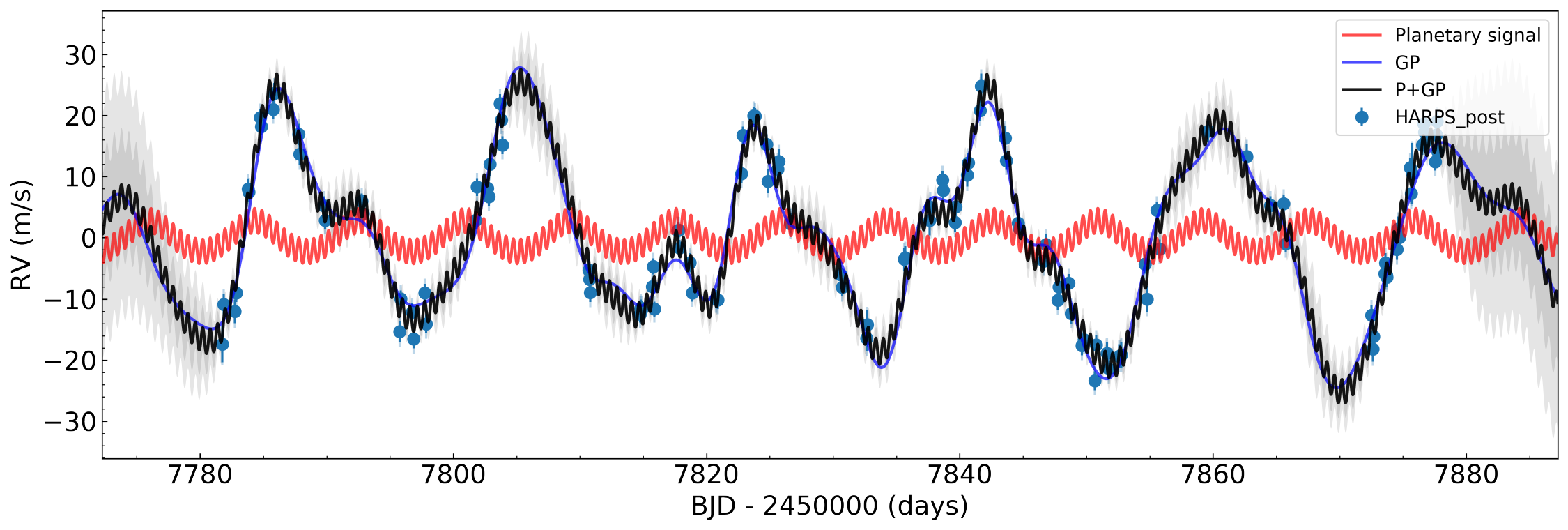}
    \caption{Best-fitting 2-planet orbital model for TIC 9870809. HARPS data shown by blue circles as a function of time. The best-fitting model for the planet signals is shown in red, the 3D GP model in blue, and the combined planets and GP shown in black. The dark and light shaded areas showing the 1$\sigma$ and 2$\sigma$ credible intervals of the corresponding GP model, respectively.}
    \label{fig:time_987}
\end{figure*}

\begin{figure}[h!]
    \centering
    \includegraphics[width=0.8\linewidth]{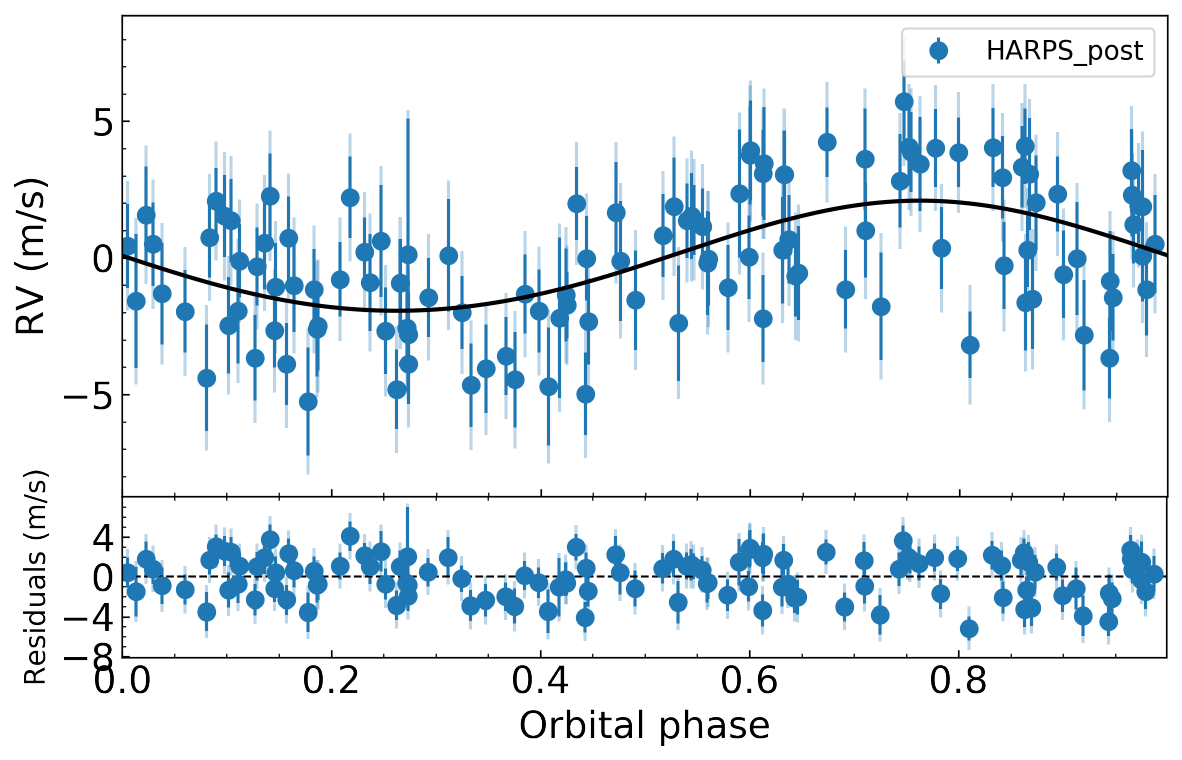}
    \caption{The phase-folded RV data from HARPS (blue circles) alongside the best-fitting planet model for TIC 9870809 b (with the effect of the other planet and the GP model subtracted). The lower part shows the residuals from the fit. There appears to be no trends visible in the residuals.}
    \label{fig:987_b}
\end{figure}

\begin{figure}[h!]
    \centering
    \includegraphics[width=0.8\linewidth]{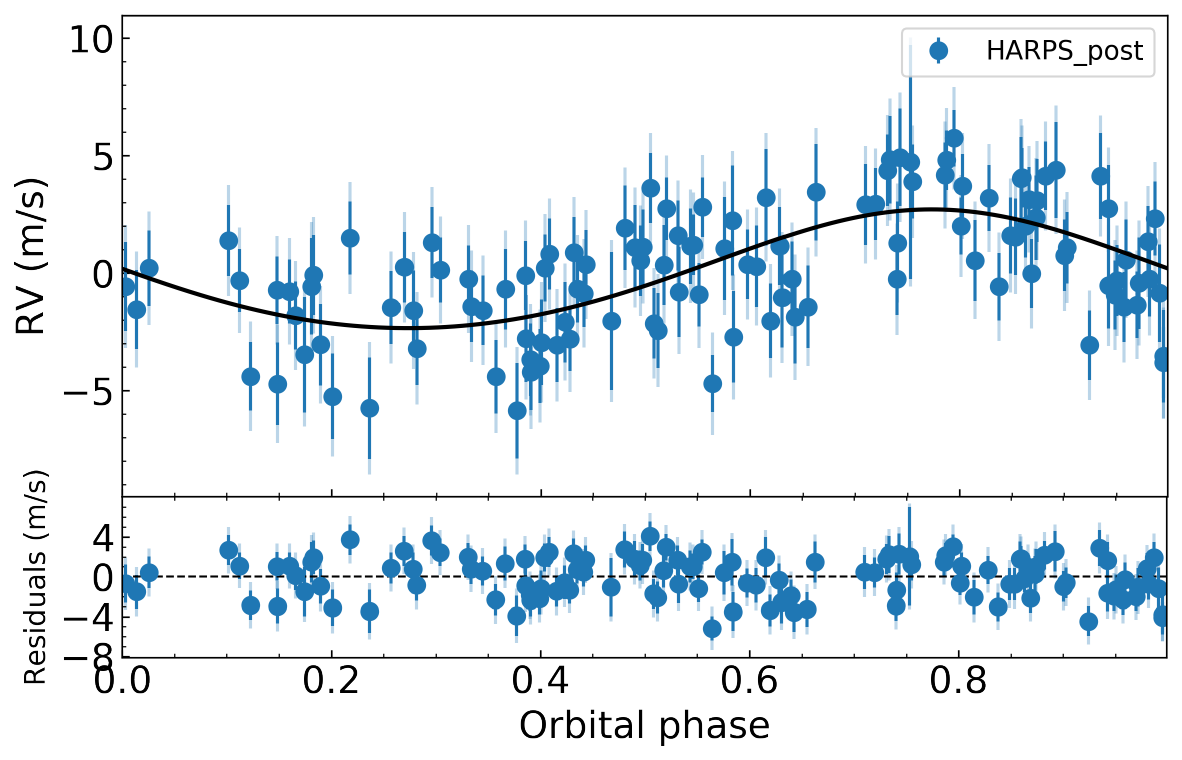}
    \caption{The phase-folded RV data from HARPS (blue circles) alongside the best-fitting planet model for TIC 9870809 c (with the effect of the other planet and the GP model subtracted). The lower part shows the residuals from the fit. There appears to be no trends visible in the residuals.}
    \label{fig:987_c}
\end{figure}

Our best model for this target is the 3D GP model with beta distribution on the eccentricity, we focus on that specific case here. After the MCMC fitting we find the parameters given in Table \ref{tab:results_987}.
In Fig. \ref{fig:time_987} we show the full time series data for this target, with the best-fit model shown for the 3D GP with beta distribution on eccentricity case. The impact of the GP is clear in this plot, the planet signal alone would not reproduce the observations well without an activity model. In Fig. \ref{fig:987_b} and \ref{fig:987_c} we show the phase-folded RV data with best-fit model (including the GP) for planets b and c, respectively. Finally, the full posterior distribution found for all fitted parameters is shown in Fig. \ref{fig:post_987}.

\begin{figure*}
    \centering
    \includegraphics[width=0.7\linewidth]{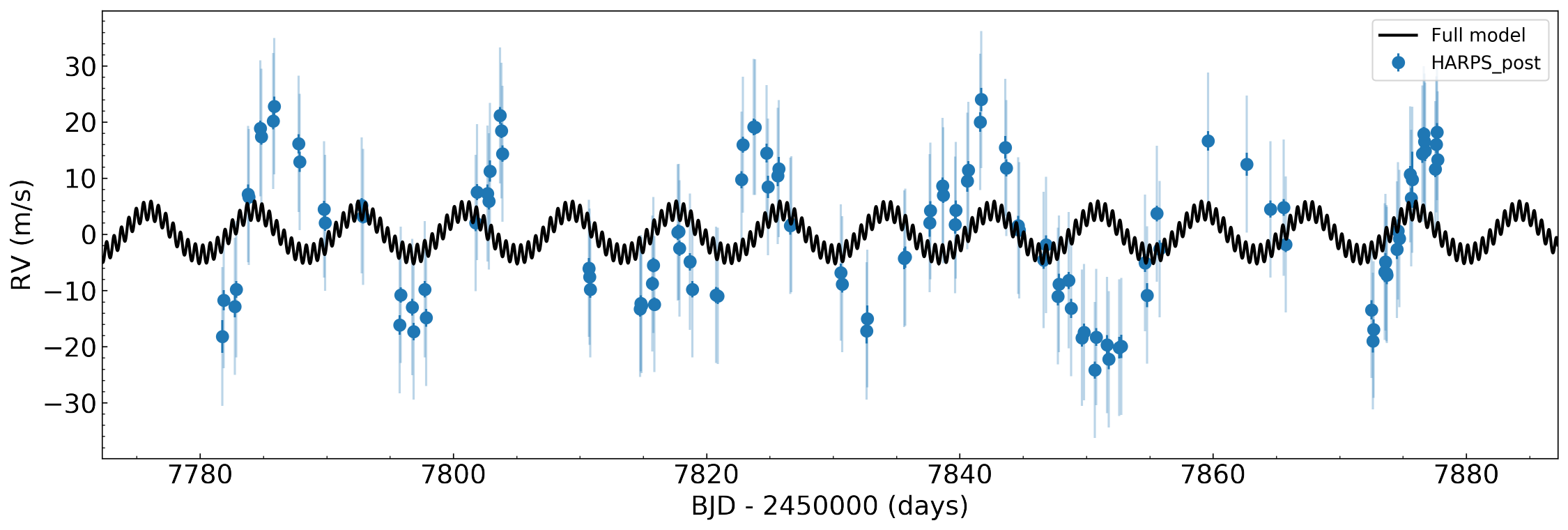}
    \caption{The no GP beta distribution on eccentricity model for TIC 9870809. HARPS data shown by blue circles as a function of time. The best-fitting model for the planet signals is shown in black. The model does not fit the data well.}
    \label{fig:time_987noGP}
\end{figure*}

\begin{figure}
    \centering
    \includegraphics[width=0.8\linewidth]{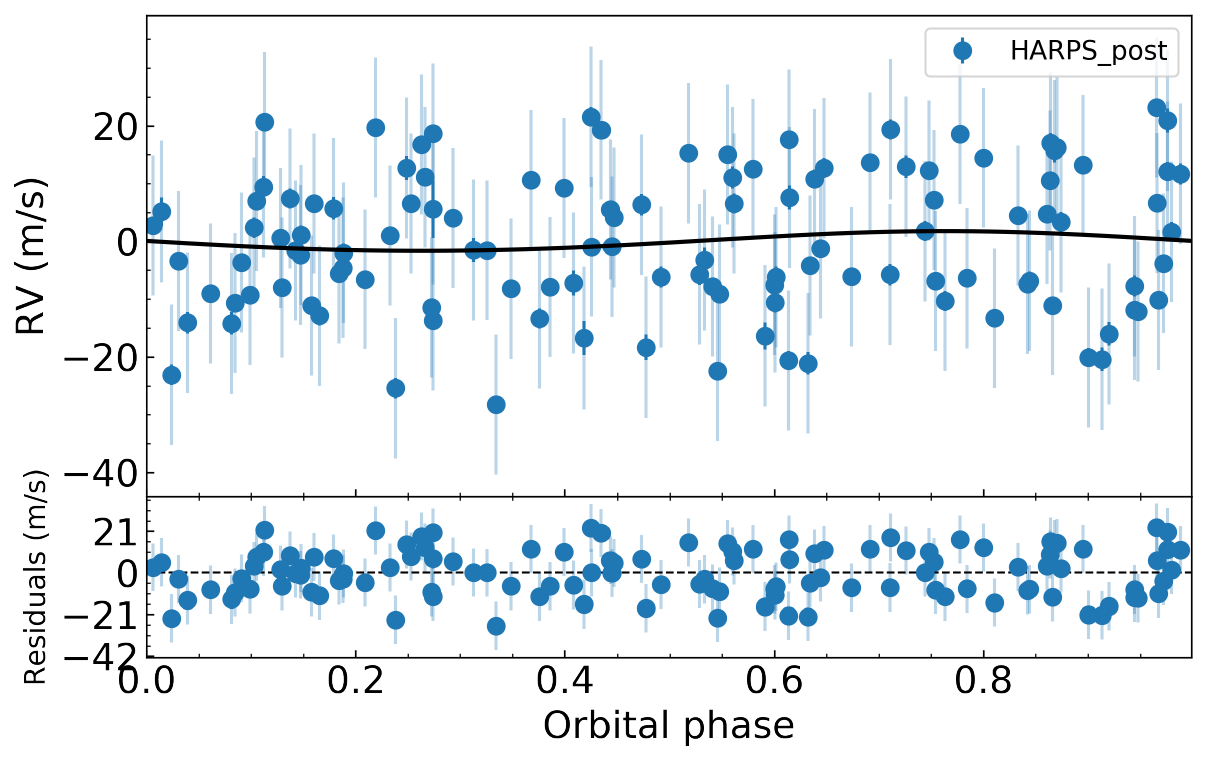}
    \caption{The phase-folded RV data from HARPS (blue circles) alongside the no GP beta distribution on eccentricity planet model for TIC 9870809 b (with the effect of the other planet subtracted). The planet signal is hardly visible.}
    \label{fig:987_bnoGP}
\end{figure}

\begin{figure}
    \centering
    \includegraphics[width=0.8\linewidth]{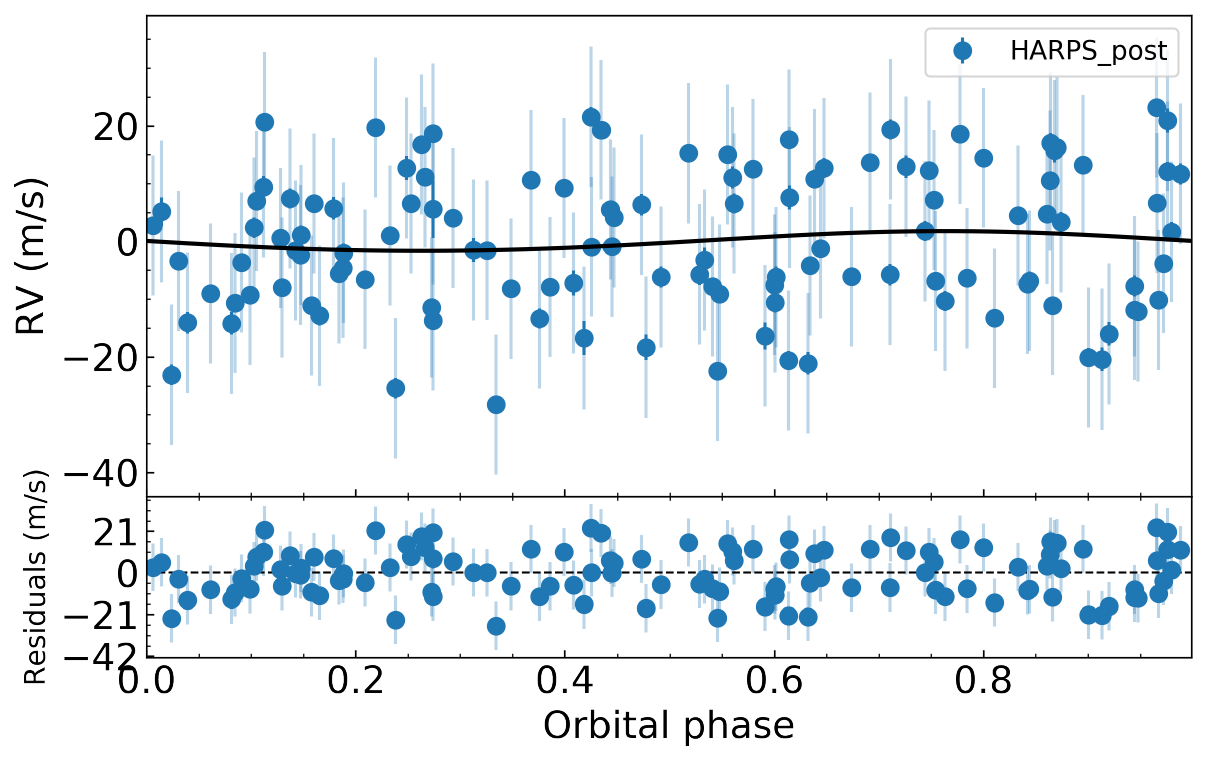}
    \caption{The phase-folded RV data from HARPS (blue circles) alongside the no GP beta distribution on eccentricity planet model for TIC 9870809 c (with the effect of the other planet subtracted). The lower part shows the residuals from the fit. The planet signal is hardly visible.}
    \label{fig:987_cnoGP}
\end{figure}

As a comparison, we now show the result plots for this same target but without a GP added to mitigate stellar activity. Specifically, the no GP model with a beta distribution on eccentricity, so we can directly compare. In Fig. \ref{fig:time_987noGP} we show the time series data with this no GP model, the data is very clearly not well fitted by this model, showing the need for the addition of a GP in this case. In Fig. \ref{fig:987_bnoGP} and Fig. \ref{fig:987_cnoGP} we show the the phase-folded RV data with best-fit model (including the GP) for planets b and c, respectively. Again, it is clear that this model does not fit the data well, and in this case the planet signals would not be recovered. Finally, in Fig. \ref{fig:post_987noGP} we show the full posterior distribution of fitted parameters for this no GP model. In this case, the RV amplitude found for both planets is not significant i.e. it is within 1$\sigma$ of 0 \ms and so using only this model with these data would result in non-detections for both planets.

\section{Caveats and recommendations}
\label{caveats}
This work has a number of challenges involved. Mostly these are related to the use of archival data. \cite{barbieri2023esoharpsradialvelocitiescatalog} discusses the possibility of some data and/or targets potentially being missing from our sample, which could impact the RV amplitudes found. We also note that we have not reprocessed the raw observations to find the RVs ourselves, this would likely reduce some of the challenges we faced and could be a useful additional step in future work.

Another difficulty with using archival data is that we have no control of the observing strategy. In some cases the cadence and baselines of observations for a target are not ideal for fitting GPs. Having long gaps between seasons of data make it harder for the GP to capture the stellar activity signal. In future we recommend teams to think about trying to reduce the length of gaps between their observing seasons where possible. We also recommend that large RV surveys are designed to mitigate the biases in observing strategy where possible, following the work of e.g. \cite{2021ApJS..256...33T}.

Another caveat of our results is that we have used only the RVs available from HARPS. For some of our targets there is additional data available from other RV instruments. In some cases this means that the published $K$ amplitudes are more precise than the ones we have found in our work. We would recommend that future work investigates the impact of adding additional data.

Finally, we note that some activity indicators or combinations of indicators may be more useful at mitigating different manifestations of stellar activity. The typical activity seen in an M dwarf star is not the same as a G type star and so having a one-size-fits-all approach may not be the most effective. However, the aim of this work is not to provide the `best' RV amplitude for each small planet, but instead to create a database of homogeneously derived planet parameters which can be used for demographics studies.

Based on the experience of this experiment, we propose a general set of recommendations for observing and modelling RVs of planet hosting stars in large batches:
\begin{enumerate}
    \item Avoid having multiple seasons of observations with large (over a year) gaps in-between if a GP is required to mitigate stellar activity signals.
    \item Avoid modelling your orbits as eccentric without a prior on the eccentricity (e.g. in the form of a beta distribution).
    \item If you use a GP, use it in combination with at least 1 activity indicator.
    \item Using a heterogeneously derived sample of planet masses will likely induce some biases when looking at a large sample: to complete demographics studies we recommend modelling planet masses in a homogeneous way.
    \item As a community we should collate a database of homogeneously derived masses wherever possible. 
\end{enumerate}

Plenty of work still needs to be done to understand the importance of homogeneous mass analysis in exoplanets. In future work it will be beneficial to look at: the impact of how the RVs are extracted from the spectra; the potential systematic difference between the RV fitting toolkits; and to which degree a joint fit with photometry changes the planet masses. We also note that it would be important to test the robustness of the RV amplitude found when changing the priors on the GP and also the choice of GP kernel, particularly for different stellar types. 

\section{Conclusions}

In this work we have refitted all publicly available HARPS radial velocity observations for 44 stars hosting small planets with planet radius smaller than 4\,$\mathrm{R_{\oplus}}$. For each target system, we have used 12 different models to investigate how model choices impact the planet mass found. 

We find that the addition of long-term trends to the model (either linear or quadratic in nature) makes a difference in some specific cases but overall there is only a very small impact from this model choice. Almost all targets have an RV amplitude within 1\,\ms of that found with a model with no trend.

The impact of eccentricity is much more significant. We find that the RV amplitudes found for fixed circular orbits differ (in some cases significantly) compared to being modelled as an eccentric orbit with wide uniform priors. We would recommend using a prior on the eccentricity, such as the beta distribution presented in \citep{VanEylen19}, to ensure that the model does not find very high values of eccentricity. 

Finally, we find that the addition of a GP, and in particular a multi-dimensional GP which fits on RVs and activity indicators, does impact the mass found. On average, the RV amplitude found is within 1\,\ms of that found for the no GP case, however this can vary up to more than 6\,\ms. The 1D GP model, fitted just on the RVs, is the most different from the others. Therefore we would recommend using either a 2D or 3D GP model for active stars. A 3D GP model does take longer computationally and so a 2D GP may be better where time is restricted and/or many targets need to be modelled.

Overall, our results demonstrate the importance of considering homogeneity in the analysis of RV observations to find planet masses. This will be particularly important for future surveys such as the PLATO mission, which aims to provide a catalogue of accurate and precise planet parameters for many new systems. To ensure this sample is accurate at the population level, it will be necessary for the analysis to be done in a homogeneous way.

In Sect.\,\ref{caveats} we describe some of the caveats of our work. In particular, we note that the RV amplitudes found in this work may not be the most precise for each individual planet in the sample. Rather, our aim is to provide a sample of masses determined homogeneously. We also note that future work should investigate the impacts of different models for different stellar types, and whether a joint model with photometric data would be of benefit. In Paper II (Osborne et al., in prep) we will investigate how the mass-radius distribution changes for our homogeneous sample, compared to the heterogeneously-derived masses. We will also comment on the future characterisation possibilities of these small planets.

\section{Data availability}

All data required to complete the analysis is publicly available on the ESO archive. However we also provide the full list of radial velocities used in our analysis at the CDS via anonymous ftp to cdsarc.u-strasbg.fr (130.79.128.5) or via \url{http://cdsweb.u-strasbg.fr/cgi-bin/qcat?J/A+A/}. The results table, Table B.2 is available at \href{https://doi.org/10.5281/zenodo.14170646}{Table B.2}.

\begin{acknowledgements}

We thank the anonymous referee for their constructive and insightful suggestions.
HLMO would like to thank the Science and
Technology Facilities Council (STFC)  and the European Southern Observatory (ESO) for funding support through PhD studentships. 
VVE has been supported by UK's Science \& Technology Facilities Council through the STFC grants ST/W001136/1 and ST/S000216/1. 
This work is based on observations collected at the European Organisation for Astronomical Research in the Southern Hemisphere under ESO programmes:
183.C-0972
192.C-0852,
60.A-9036,
072.C-0488,
198.C-0836,
091.C-0936,
1102.C-0923,
1102.C-0249,
60.A-9700,
0102.C-0584,
096.C-0499,
106.21TJ,
0100.D-0444,
188.C-0265,
198.C-0169,
0100.C-0750,
098.C-0304,
098.C-0739,
0102.C-0451,
0101.C-0510,
192.C-0224,
0103.C-0759,
0102.D-0483,
0102.C-0525,
196.C-1006,
085.C-0019,
087.C-0831,
096.C-0460,
1102.C-0249,
1102.C-0249,
60.A-9709,
1102.C-0923,
0103.C-0442,
0103.C-0874,
1102.C-0339,
183.C-0437,
1102.C-0339,
082.C-0718,
0100.C-0808,
099.C-0491,
0101.C-0788,
0103.C-0548,
0102.C-0338,
0101.C-0829,
0104.C-0413,
105.20N0,
198.C-0838,
191.C-0873,
095.C-0718,
0100.C-0884,
283.C-5022,
098.C-0860,
097.C-0948,
074.C-0364,
0101.C-0497,
0101.C-0407,
282.C-5036,
082.C-0308, and 
088.C-0323. 
All data is publicly available on the ESO archive.
We thank the PIs of the programmes: 
Udry, 
Mayor, 
Diaz, 
Gandolfi, 
Armstrong, 
Demedeiros, 
Lorenzo-Oliveira, 
Melendez, 
Santerne, 
Ehrenreich, 
Lagrange, 
Benatti, 
Berdinas, 
Lo Curto, 
Nielsen, 
Bonfils, 
Allart, 
Trifonov, 
Brahm, 
Albrecht, 
Astudillo, 
NIRPS GTO Team, and 
Robichon.
We also thank the many people involved in the observing and archiving of the data, without whom this work would not have been possible.

\end{acknowledgements}

\bibliographystyle{aa}
\bibliography{thebib}

%
%
\appendix

\section{Individual systems}
We aimed in all cases to treat the data homogeneously for every target so that the same steps could be applied to everything automatically. However, in a few cases we noticed some issues with individual systems which required us to make manual changes to the input files for those. We tried to keep everything else homogeneous in our analysis, and in most cases it just involved excluding some of the RV data points. These specific cases are detailed below.

\subsection{Removal of RV data}
\label{removal_rvs}
For the system TIC\,173103335, we noticed that quite a bit of RV data was largely offset (more than 30km/s) from the rest of the RVs. If we include this data in the fit then the model struggles to find a solution, particularly in the cases including a GP where the additional GP hyperparameters allow for possible over fitting. For this reason, we removed the largely outlying data from this system (we cut out all data with RV values <10km/s. As this is archival data it is difficult to know why in some cases there would be such a large offset; we believe it is likely due to the incorrect stellar mask being applied for the CCF data reduction in the HARPS pipeline.

For TIC\,220479565, we again see that this system appears to have some largely offset RV data which makes the modelling very challenging. We choose to cut all data where the RV is negative (i.e. we cut at RV = 0 km/s). 
In the case of TIC\,260004324 we also cut the outlying data points at the threshold of 42.5 km/s. This successfully removes the largely outlying RV data. 
TIC\,56815340 has a large amount of in-transit RV observations, this means that many exposures are taken over the course of one observing night. Because of this, the fitting with a GP takes much longer and can be confused by the many points over one night. For this reason, we remove any data for this target where there are more than 5 observations in a single date. 

\subsection{Defining priors}
\label{manualpriors}
For the majority of our targets, the NASA Exoplanet Archive provides details of the orbital period and time of mid-transit, $T_c$ (or time of inferior conjunction for non-transiting companions). However, a few systems do not have $T_c$ listed and so for these we search the published literature for these planets and manually input the values. Table \ref{tab:manual_priors} gives the specific values for these priors and the reference they were taken from. For one system, TIC\,280304863, there was no $T_c$ given for planet d, so for this planet we set a wide Gaussian priors on the $T_c$ for planet c.

\begin{table}
    \caption{The $T_c$ values added manually for targets where the NASA Exoplanet Archive does not provide these.}
    \begin{tabular}{clcc}
    \hline
        TIC ID & & $T_c$ (-2450000) &  Reference \\
    \hline
         307210830 & e & 8439.40 $\pm$ 0.37 &  \protect{\cite{2021A&A...653A..41D}}\\
         388804061 & c & 7264.55 $\pm$ 0.46 &  \protect{\cite{2017A&A...608A..35C}}\\
         101955023 & c & 7506.02 $\pm$ 0.34  &  \protect{\cite{2018A&A...618A.142B}}\\
         413248763 & c & 8314.30 $\pm$ 0.42  &  \protect{\cite{2019A&A...628A..39L}}\\
         413248763 & d & 8326.10 $\pm$ 3.9 &  \protect{\cite{2019A&A...628A..39L}}\\
         299799658 & c & 9087.61 $\pm$ 1.84  &  \protect{\cite{2021A&A...653A.105O}}\\
         73228647  & c & 8798.17 $\pm$ 0.19 &  \protect{\cite{2021MNRAS.502.4842O}}\\
         280304863 & d$^a$ & 4445.00 $\pm$ 20 &  \protect{\cite{2009A&A...506..303Q}}\\
    \hline \\
    \end{tabular}
    \\
    $a$: For this planet no papers provide a $T_0$ value and so we use the value for planet c, in the listed reference, with a wide uncertainty of $\sim$2 times the orbital period.
    \label{tab:manual_priors}
\end{table}

\begin{table}[]
\centering
\caption{Priors used for the 3D GP model with beta distribution on eccentricity for TIC 9870809.}
\label{tab:priors_987}
\begin{tabular}{@{}ll@{}}
\toprule

Planet b Priors \\
\hline
T0 & = g{[} 7646.5666 , 0.0005 {]} \\ 
P & = g{[} 0.5842 , 0.0000 {]} \\
e & = b{[} 1.5200 , 29.0000 {]} \\
w & = f{[} 0.0000 , 6.2832 {]} \\
b & = f{[} 0.0000 , 0.0000 {]} \\
a & = f{[} 1.5000 , 1000.0000 {]} \\
rp & = f{[} 0.0000 , 0.0000 {]} \\
K & = u{[} 0.0000 , 0.5000 {]}  \\
 \hline
 Planet c Priors \\
 \hline
T0 & = g{[} 7586.3431 , 0.0019 {]} \\
P & = g{[} 8.3273 , 0.0004 {]} \\
e & = b{[} 1.5200 , 29.0000 {]} \\
w & = f{[} 0.0000 , 6.2832 {]} \\
b & = f{[} 0.0000 , 0.0000 {]} \\
a & = f{[} 1.5000 , 1000.0000 {]} \\
rp & = f{[} 0.0000 , 0.0000 {]} \\
K & = u{[} 0.0000 , 0.5000 {]} \\
 \hline
 Other Parameter Priors \\
 \hline
q1 & = f{[} 0.0000 , 1.0000 {]} \\
q2 & = f{[} 0.0000 , 1.0000 {]} \\
HARPS\_post &  = u{[} 22.4567 , 23.5048 {]} \\
FWHM\_post &  = u{[} 6.6676 , 7.7970 {]} \\
BIS\_post &  = u{[} -0.5328 , 0.5252 {]} \\ \bottomrule
\end{tabular}
\\
\footnotesize{The priors are given as $g$, normal distribution, $f$, fixed value, $b$, beta distribution, or $u$, uniform distribution.}
\end{table}

\begin{figure}
    \centering
    \includegraphics[width=\linewidth]{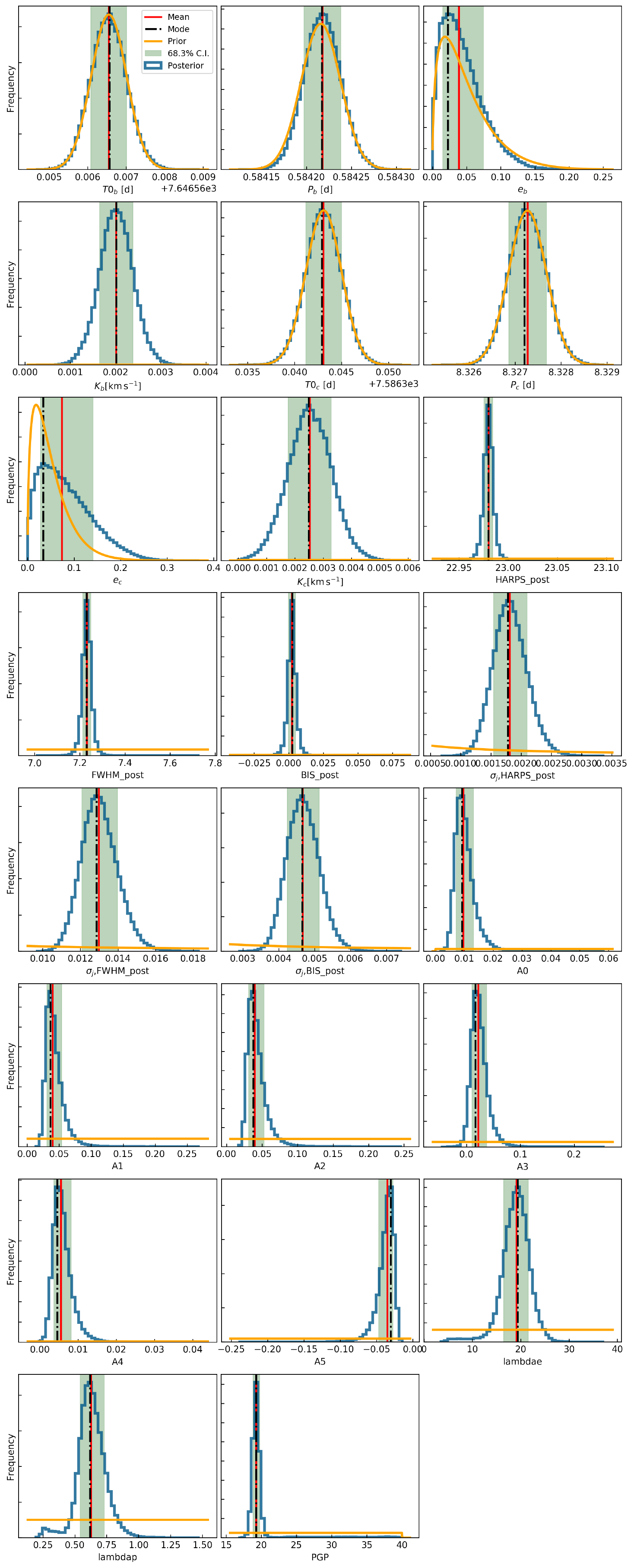}
    \caption{Full posterior distribution of fitted parameters for TIC 98720809 with the 3D GP beta distribution model.}
    \label{fig:post_987}
\end{figure}

\begin{figure}
    \centering
    \includegraphics[width=\linewidth]{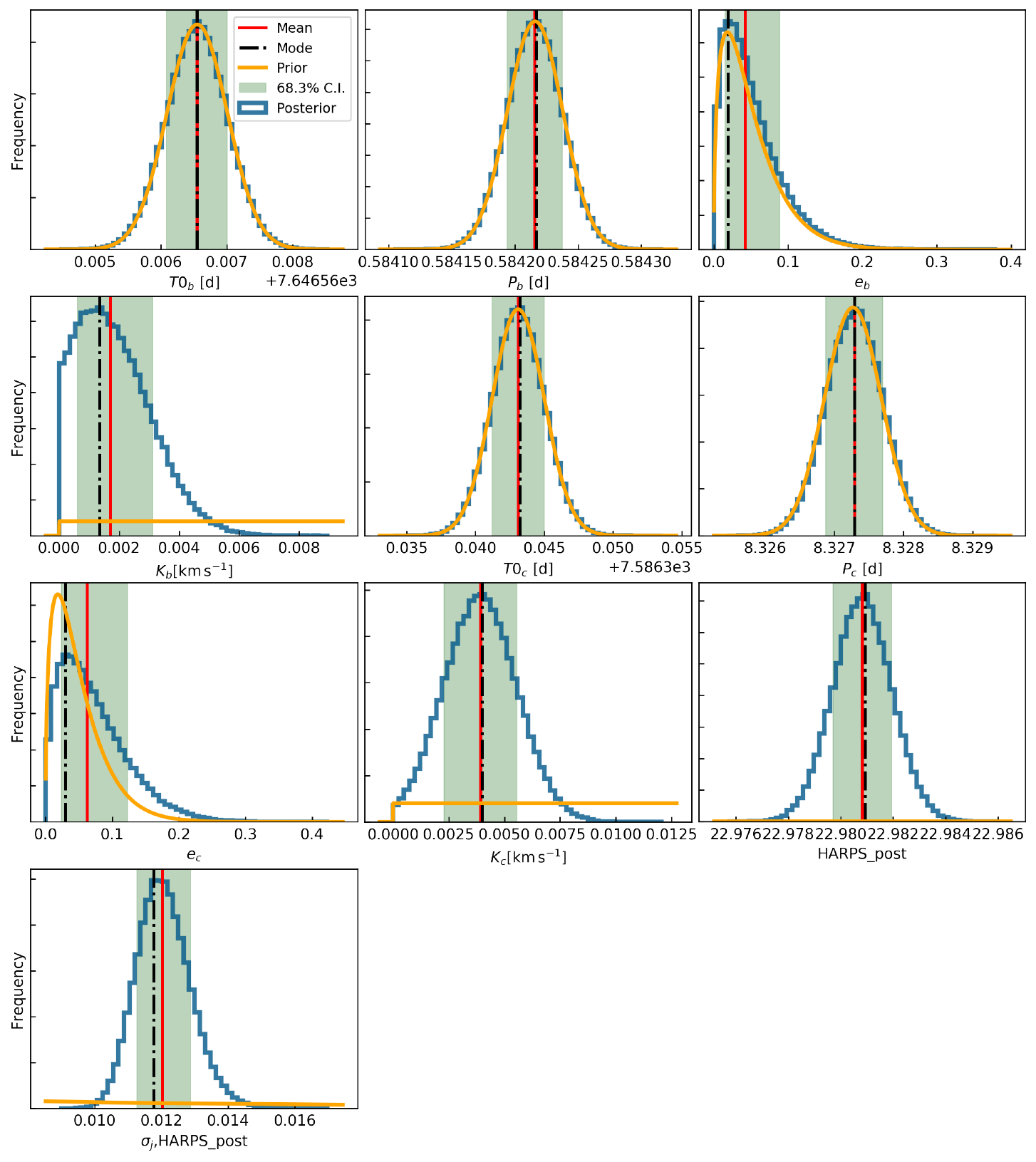}
    \caption{Full posterior distribution of fitted parameters for TIC 98720809 with the no GP beta distribution model.}
    \label{fig:post_987noGP}
\end{figure}

\begin{table*}[]
\centering
\caption{Fitted and derived parameters for the 2-planet system TIC 9870809.}
\label{tab:results_987}
\begin{tabular}{@{}ll@{}}
\toprule

Output Summary  &\\
\hline
chi2            & = 342.0834 \\ 
dof &     = 337 \\
chi2/dof         &= 1.0151 \\
ln likelihood    &= 1289.9751 \\
BIC              &= -2444.5698 \\
AIC              &= -2533.9501 \\
\hline
Planet b & Fitted\\
\hline

T0       &  = 7646.5665430 - 0.0004630 + 0.0004680 days \\
P         & = 0.5842176 - 0.0000205 + 0.0000206 days \\
e         & = 0.0387125 - 0.0238209 + 0.0360824 \\
w         & = 0.0000000 - 0.0000000 + 0.0000000 deg \\
K         & = 2.0147679 - 0.3621305 + 0.3688024 m/s \\
\hline
 & Derived\\
\hline
Mp         &= 2.2259962 - 0.3999928 + 0.4080294 M\_earth \\
Tperi      &= 7646.4276887 - 0.0044212 + 0.0067113 days \\
\hline
Planet c & Fitted\\
\hline
T0      &   = 7586.3430800 - 0.0018820 + 0.0019140 days \\
P        &  = 8.3272693 - 0.0004023 + 0.0004133 days \\
e         & = 0.0734197 - 0.0459481 + 0.0672700 \\
w          &= 0.0000000 - 0.0000000 + 0.0000000 deg \\
K         & = 2.5267559 - 0.7660638 + 0.7617282 m/s \\
\hline
& Derived \\
\hline
Mp         &= 6.7427960 - 2.0410651 + 2.0289184 M\_earth \\
Tperi      &= 7584.4557591 - 0.1217128 + 0.1772512 days \\
\hline
Other parameters & \\
\hline
Sys. vel. HARPS\_post &= 22.9800205 - 0.0045698 + 0.0044317 km/s \\
Sys. vel. FWHM\_post &= 7.2315287 - 0.0189790 + 0.0185481 km/s \\
Sys. vel. BIS\_post &= 0.0023307 - 0.0026849 + 0.0026301 km/s \\
\hline
HARPS\_post jitter &= 1.8102796 - 0.2667877 + 0.2901762 m/s \\
FWHM\_post jitter &= 12.9766515 - 0.8945287 + 1.0032974 m/s \\
BIS\_post jitter &= 4.6689670 - 0.4317737 + 0.4590220 m/s \\
\hline
A0   &      = 0.0096250 - 0.0024629 + 0.0035341 \\
A1    &     = 0.0398090 - 0.0088824 + 0.0142108 \\
A2     &    = 0.0396643 - 0.0085181 + 0.0133253 \\
A3      &   = 0.0219536 - 0.0118040 + 0.0156001 \\
A4       &  = 0.0055228 - 0.0018464 + 0.0026643 \\
A5        & = -0.0350163 - 0.0119756 + 0.0076725 \\
lambdae    &= 19.1165742 - 2.6449472 + 2.4364964 \\
lambdap    &= 0.6272393 - 0.0863693 + 0.1037217 \\
PGP        &= 19.2901787 - 0.4651249 + 0.5123584 \\ \bottomrule
\end{tabular}
\end{table*}

\onecolumn
\begin{landscape}
\section{Fitted parameters for our chosen best models}

\begin{tiny}
\begin{longtable}[c]{@{}lllllllllllllll@{}}

\caption{Fitted parameters for the best model chosen for each planet.}
\label{best_model_table}\\
\toprule
TIC ID & Planet & Model & K (m/s) & -eK (m/s) & +eK (m/s) & Eccen & -eEccen & +eEccen & Period (days) & -ePeriod (days) & +ePeriod (days) & Mass M$\oplus$& -eMass M$\oplus$& +eMass M$\oplus$\\* \midrule
\endfirsthead
\multicolumn{15}{c}%
{{\bfseries Table \thetable\ continued from previous page}} \\
\toprule
TIC ID & Planet & Model & K (m/s) & -eK (m/s) & +eK (m/s) & Eccen & -eEccen & +eEccen & Period (days) & -ePeriod (days) & +ePeriod (days) & Mass M$\oplus$& -eMass M$\oplus$& +eMass M$\oplus$\\* \midrule
\endhead
\bottomrule
\endfoot
\endlastfoot
280304863 & b & n & 4.61 & 0.63 & 0.64 & 0.02 & 0.01 & 0.03 & 0.85 & 0.00E+00 & 0.00E+00 & 5.91 & 1.12 & 1.22 \\
437444661 & c & n & 1.66 & 1.01 & 1.11 & 0.05 & 0.03 & 0.05 & 20.66 & 7.89E-04 & 7.58E-04 & 6.28 & 3.82 & 4.20 \\
437444661 & d & n & 0.67 & 0.48 & 0.88 & 0.04 & 0.03 & 0.05 & 31.72 & 1.07E-03 & 1.02E-03 & 2.92 & 2.09 & 3.83 \\
437444661 & e & n & 0.77 & 0.48 & 0.68 & 0.05 & 0.03 & 0.05 & 4.35 & 3.82E-04 & 3.73E-04 & 1.73 & 1.08 & 1.54 \\
101955023 & b & n & 2.73 & 0.45 & 0.45 & 0.03 & 0.02 & 0.03 & 1.63 & 2.71e-05 & 2.61e-05 & 1.59 & 0.28 & 0.30 \\
467929202 & b & e & 13.73 & 0.73 & 0.73 & 0.02 & 0.01 & 0.02 & 1.58 & 0.00E+00 & 0.00E+00 & 13.01 & 1.81 & 1.79 \\
56815340 & c & e & 2.77 & 0.99 & 1.02 & 0.04 & 0.03 & 0.04 & 21.06 & 4.57E-04 & 4.61E-04 & 13.83 & 4.93 & 5.08 \\
56815340 & b & e & 2.84 & 1.12 & 1.16 & 0.06 & 0.04 & 0.05 & 9.55 & 8.81E-04 & 8.89E-04 & 10.87 & 4.27 & 4.43 \\
136916387 & b & n & 1.55 & 0.12 & 0.13 & 0.02 & 0.01 & 0.02 & 11.58 & 7.7e-05 & 7.54e-05 & 4.98 & 0.43 & 0.44 \\
136916387 & c & n & 2.11 & 0.14 & 0.15 & 0.02 & 0.01 & 0.02 & 27.59 & 1.08E-04 & 1.07E-04 & 9.06 & 0.67 & 0.70 \\
136916387 & d & n & 1.43 & 0.15 & 0.14 & 0.03 & 0.02 & 0.03 & 107.25 & 3.80E-02 & 3.91E-02 & 9.67 & 1.05 & 1.00 \\
388804061 & b & e & 2.51 & 0.72 & 0.71 & 0.07 & 0.04 & 0.06 & 32.94 & 9.62e-05 & 9.58e-05 & 7.27 & 2.10 & 2.09 \\
173103335 & b & e & 2.60 & 0.90 & 0.90 & 0.03 & 0.02 & 0.04 & 10.05 & 8.9e-06 & 8.8e-06 & 5.51 & 1.91 & 1.93 \\
173103335 & c & e & 1.04 & 0.64 & 0.78 & 0.04 & 0.03 & 0.05 & 24.65 & 4.33e-05 & 4.4e-05 & 2.98 & 1.82 & 2.24 \\
173103335 & d & e & 0.59 & 0.42 & 0.72 & 0.04 & 0.02 & 0.04 & 44.56 & 1.32E-04 & 1.30E-04 & 2.04 & 1.47 & 2.51 \\
92226327 & b & n & 5.37 & 0.58 & 0.66 & 0.05 & 0.03 & 0.04 & 24.74 & 1.98e-05 & 1.95e-05 & 7.78 & 0.84 & 0.95 \\
92226327 & c & n & 2.00 & 0.42 & 0.41 & 0.04 & 0.03 & 0.04 & 3.78 & 2e-06 & 2e-06 & 1.55 & 0.32 & 0.32 \\
4610830 & e & n & 3.65 & 0.72 & 0.72 & 0.03 & 0.02 & 0.03 & 8.26 & 1.98E-04 & 1.97E-04 & 11.07 & 2.17 & 2.20 \\
4610830 & d & n & 3.17 & 0.61 & 0.63 & 0.03 & 0.02 & 0.03 & 5.40 & 2.07E-04 & 2.04E-04 & 8.33 & 1.59 & 1.68 \\
4610830 & f & n & 0.38 & 0.28 & 0.52 & 0.04 & 0.03 & 0.05 & 12.76 & 5.02E-04 & 4.93E-04 & 1.34 & 0.96 & 1.80 \\
4610830 & b & n & 1.71 & 0.59 & 0.56 & 0.03 & 0.02 & 0.04 & 2.35 & 2.13E-04 & 2.07E-04 & 3.41 & 1.18 & 1.13 \\
4610830 & c & n & 2.52 & 0.58 & 0.56 & 0.03 & 0.02 & 0.03 & 3.56 & 1.21E-04 & 1.23E-04 & 5.77 & 1.32 & 1.29 \\
4610830 & g & n & 1.27 & 0.73 & 0.87 & 0.05 & 0.03 & 0.06 & 41.97 & 8.66E-03 & 8.61E-03 & 6.59 & 3.81 & 4.52 \\
428820090 & b & n & 1.18 & 0.74 & 0.86 & 0.04 & 0.03 & 0.05 & 2.47 & 3.96e-05 & 4.03e-05 & 2.13 & 1.33 & 1.55 \\
428820090 & c & n & 2.72 & 0.78 & 0.76 & 0.04 & 0.03 & 0.04 & 7.06 & 2.38E-04 & 2.45E-04 & 6.96 & 1.99 & 1.96 \\
428820090 & d & n & 1.22 & 0.67 & 0.71 & 0.03 & 0.02 & 0.03 & 24.36 & 6.85E-04 & 6.57E-04 & 4.71 & 2.58 & 2.74 \\
374180079 & c & n & 0.51 & 0.34 & 0.45 & 0.05 & 0.03 & 0.05 & 7.81 & 1.88E-03 & 1.91E-03 & 1.24 & 0.82 & 1.10 \\
374180079 & e & n & 3.22 & 0.57 & 0.54 & 0.02 & 0.01 & 0.02 & 19.48 & 1.14E-03 & 1.16E-03 & 10.55 & 1.89 & 1.83 \\
374180079 & d & n & 2.56 & 0.51 & 0.49 & 0.05 & 0.03 & 0.05 & 14.70 & 3.43E-04 & 3.35E-04 & 7.64 & 1.53 & 1.50 \\
374180079 & b & n & 2.85 & 0.40 & 0.39 & 0.03 & 0.02 & 0.03 & 0.66 & 1.5e-05 & 1.54e-05 & 3.02 & 0.43 & 0.43 \\
146364192 & b & n & 3.62 & 0.46 & 0.48 & 0.03 & 0.02 & 0.03 & 2.37 & 7.33e-05 & 7.38e-05 & 7.13 & 0.91 & 0.95 \\
98720809 & b & n & 2.01 & 0.36 & 0.37 & 0.04 & 0.02 & 0.04 & 0.58 & 2.05e-05 & 2.06e-05 & 2.40 & 0.43 & 0.44 \\
98720809 & c & n & 2.53 & 0.77 & 0.76 & 0.07 & 0.05 & 0.07 & 8.33 & 4.02E-04 & 4.13E-04 & 7.26 & 2.20 & 2.18 \\
279741379 & b & n & 4.49 & 0.59 & 0.57 & 0.04 & 0.02 & 0.03 & 35.61 & 5.89E-04 & 5.67E-04 & 18.60 & 2.64 & 2.74 \\
52368076 & b & n & 3.49 & 0.47 & 0.48 & 0.09 & 0.04 & 0.04 & 4.65 & 3.30E-04 & 3.35E-04 & 8.20 & 1.13 & 1.16 \\
52368076 & c & n & 2.17 & 0.48 & 0.47 & 0.03 & 0.02 & 0.03 & 9.15 & 6.97E-04 & 7.17E-04 & 6.40 & 1.43 & 1.40 \\
52368076 & d & n & 2.73 & 0.82 & 0.81 & 0.06 & 0.04 & 0.06 & 19.98 & 4.95E-03 & 5.00E-03 & 10.45 & 3.12 & 3.10 \\
259377017 & b & e & 1.51 & 0.61 & 0.62 & 0.05 & 0.03 & 0.05 & 3.36 & 4.8e-06 & 4.8e-06 & 1.94 & 0.78 & 0.80 \\
259377017 & c & e & 3.02 & 0.61 & 0.61 & 0.03 & 0.02 & 0.04 & 5.66 & 3.1e-06 & 3.1e-06 & 4.63 & 0.95 & 0.96 \\
259377017 & d & e & 2.84 & 0.62 & 0.63 & 0.04 & 0.03 & 0.04 & 11.38 & 1.31e-05 & 1.31e-05 & 5.50 & 1.22 & 1.24 \\
307210830 & b & n & 0.20 & 0.14 & 0.25 & 0.04 & 0.02 & 0.04 & 2.25 & 1.1e-06 & 1.2e-06 & 0.17 & 0.12 & 0.21 \\
307210830 & c & n & 1.93 & 0.33 & 0.31 & 0.03 & 0.02 & 0.03 & 3.69 & 1.5e-06 & 1.5e-06 & 1.94 & 0.35 & 0.35 \\
307210830 & d & n & 1.08 & 0.33 & 0.31 & 0.03 & 0.02 & 0.04 & 7.45 & 8.6e-06 & 8.3e-06 & 1.37 & 0.43 & 0.42 \\
413248763 & b & e & 1.52 & 0.53 & 0.52 & 0.04 & 0.03 & 0.04 & 3.93 & 2e-06 & 2e-06 & 1.99 & 0.69 & 0.69 \\
178155732 & b & e & 1.24 & 0.61 & 0.66 & 0.03 & 0.02 & 0.04 & 3.59 & 1.52E-04 & 1.52E-04 & 3.34 & 1.66 & 1.78 \\
178155732 & c & e & 0.48 & 0.33 & 0.51 & 0.04 & 0.03 & 0.04 & 5.97 & 5.94E-04 & 6.01E-04 & 1.52 & 1.05 & 1.62 \\
178155732 & d & e & 1.07 & 0.60 & 0.66 & 0.04 & 0.03 & 0.05 & 11.23 & 1.11E-03 & 1.09E-03 & 4.22 & 2.37 & 2.62 \\
150428135 & e & e & 0.59 & 0.40 & 0.59 & 0.04 & 0.02 & 0.04 & 27.81 & 4.60E-04 & 4.56E-04 & 1.56 & 1.06 & 1.56 \\
150428135 & c & e & 0.43 & 0.31 & 0.51 & 0.04 & 0.02 & 0.04 & 16.05 & 1.98e-05 & 2.02e-05 & 0.94 & 0.67 & 1.12 \\
150428135 & d & e & 0.46 & 0.33 & 0.58 & 0.03 & 0.02 & 0.04 & 37.42 & 3.93E-04 & 3.95E-04 & 1.35 & 0.98 & 1.72 \\
150428135 & b & e & 1.39 & 0.63 & 0.65 & 0.04 & 0.02 & 0.04 & 9.98 & 4.16e-05 & 4.04e-05 & 2.61 & 1.18 & 1.24 \\
432254760 & c & e & 3.57 & 0.57 & 0.57 & 0.04 & 0.02 & 0.03 & 15.62 & 9.87E-04 & 9.92E-04 & 14.79 & 2.37 & 2.36 \\
432254760 & b & e & 4.12 & 0.59 & 0.57 & 0.03 & 0.02 & 0.03 & 3.60 & 2.62E-04 & 2.63E-04 & 10.47 & 1.50 & 1.46 \\
432254760 & d & e & 1.54 & 0.55 & 0.54 & 0.03 & 0.02 & 0.04 & 35.75 & 5.02E-03 & 4.96E-03 & 8.43 & 3.00 & 2.98 \\
260004324 & b & e & 0.67 & 0.43 & 0.57 & 0.04 & 0.03 & 0.04 & 3.81 & 8.32e-05 & 8.43e-05 & 1.00 & 0.65 & 0.85 \\
254113311 & b & n & 6.97 & 1.58 & 1.66 & 0.03 & 0.02 & 0.03 & 4.07 & 4.99E-04 & 4.60E-04 & 13.84 & 3.15 & 3.32 \\
260647166 & e & n & 1.67 & 0.31 & 0.31 & 0.05 & 0.03 & 0.04 & 19.59 & 8.45e-05 & 8.28e-05 & 6.37 & 1.21 & 1.22 \\
260647166 & c & n & 1.56 & 0.24 & 0.24 & 0.08 & 0.04 & 0.05 & 6.20 & 4.88e-05 & 5.13e-05 & 4.07 & 0.63 & 0.64 \\
260647166 & d & n & 1.15 & 0.27 & 0.27 & 0.05 & 0.03 & 0.05 & 14.18 & 8.4e-05 & 8.32e-05 & 3.95 & 0.95 & 0.96 \\
260647166 & b & n & 2.14 & 0.22 & 0.22 & 0.03 & 0.02 & 0.03 & 3.80 & 2.48e-05 & 2.42e-05 & 4.76 & 0.52 & 0.52 \\
260647166 & f & n & 0.89 & 0.33 & 0.33 & 0.03 & 0.02 & 0.04 & 29.54 & 3.28E-04 & 3.34E-04 & 3.90 & 1.46 & 1.46 \\
94986319 & b & n & 2.73 & 0.30 & 0.30 & 0.02 & 0.01 & 0.02 & 5.20 & 4.60E-04 & 4.60E-04 & 6.63 & 0.75 & 0.75 \\
178819686 & b & n & 3.77 & 0.30 & 0.31 & 0.05 & 0.02 & 0.03 & 5.61 & 1.23E-03 & 1.20E-03 & 9.89 & 0.82 & 0.84 \\
178819686 & c & n & 2.08 & 0.56 & 0.65 & 0.04 & 0.02 & 0.04 & 12.27 & 5.06E-03 & 5.05E-03 & 7.07 & 1.91 & 2.21 \\
306996324 & c & e & 1.96 & 0.87 & 0.91 & 0.09 & 0.05 & 0.07 & 15.67 & 7.52e-05 & 7.5e-05 & 4.71 & 2.08 & 2.18 \\
306996324 & b & e & 2.13 & 0.74 & 0.75 & 0.04 & 0.02 & 0.04 & 8.25 & 2.4e-05 & 2.39e-05 & 4.14 & 1.45 & 1.47 \\
73228647 & b & n & 2.32 & 0.32 & 0.32 & 0.04 & 0.03 & 0.04 & 2.54 & 3.93E-04 & 3.87E-04 & 4.94 & 0.70 & 0.73 \\
73228647 & c & n & 3.23 & 0.53 & 0.35 & 0.02 & 0.01 & 0.03 & 6.73 & 3.58E-03 & 3.97E-03 & 9.49 & 1.61 & 1.12 \\
320004517 & b & n & 3.09 & 1.23 & 1.18 & 0.04 & 0.03 & 0.05 & 17.47 & 5.54e-05 & 5.81e-05 & 12.75 & 5.08 & 4.90 \\
34068865 & b & n & 0.82 & 0.15 & 0.14 & 0.03 & 0.02 & 0.03 & 0.32 & 2e-07 & 2e-07 & 0.52 & 0.10 & 0.09 \\
346250894 & b & e & 4.73 & 2.17 & 2.24 & 0.04 & 0.03 & 0.05 & 1.62 & 7.54e-05 & 7.5e-05 & 9.18 & 4.21 & 4.34 \\
101011575 & b & n & 2.52 & 1.07 & 1.05 & 0.03 & 0.02 & 0.03 & 6.40 & 6.8e-06 & 6.5e-06 & 5.91 & 2.51 & 2.48 \\
101011575 & c & n & 3.19 & 0.72 & 0.84 & 0.06 & 0.03 & 0.05 & 18.88 & 8.27E-04 & 8.41E-04 & 10.75 & 2.44 & 2.84 \\
299799658 & b & n & 3.13 & 0.61 & 0.78 & 0.03 & 0.02 & 0.03 & 4.11 & 1.39E-03 & 1.32E-03 & 7.52 & 1.48 & 1.87 \\
150098860 & b & n & 4.41 & 0.57 & 0.55 & 0.02 & 0.01 & 0.02 & 10.70 & 8.66e-05 & 8.58e-05 & 13.41 & 1.75 & 1.71 \\
31374837 & b & n & 3.48 & 0.39 & 0.39 & 0.02 & 0.01 & 0.02 & 0.49 & 2.9e-06 & 2.6e-06 & 3.62 & 0.45 & 0.47 \\
31374837 & d & n & 3.22 & 0.32 & 0.34 & 0.03 & 0.02 & 0.03 & 12.46 & 2.05e-05 & 1.95e-05 & 9.86 & 1.13 & 1.21 \\
134200185 & b & n & 1.29 & 0.17 & 0.18 & 0.04 & 0.02 & 0.04 & 0.55 & 1.45e-05 & 1.39e-05 & 1.35 & 0.18 & 0.19 \\
50618703 & b & n & 2.37 & 0.55 & 0.56 & 0.05 & 0.03 & 0.05 & 1.55 & 2e-06 & 2e-06 & 3.14 & 0.73 & 0.74 \\
262530407 & b & n & 3.20 & 0.72 & 0.79 & 0.03 & 0.02 & 0.04 & 2.85 & 2.3e-06 & 2.3e-06 & 4.59 & 1.04 & 1.13 \\
440887364 & b & n & 2.76 & 0.47 & 0.49 & 0.06 & 0.03 & 0.05 & 3.82 & 9.9e-06 & 9.8e-06 & 5.19 & 0.92 & 0.97 \\
440887364 & c & n & 3.08 & 0.56 & 0.53 & 0.02 & 0.01 & 0.02 & 8.60 & 9.9e-06 & 9.9e-06 & 7.61 & 1.41 & 1.40 \\
317060587 & b & e & 4.57 & 0.57 & 0.57 & 0.01 & 0.01 & 0.02 & 9.14 & 1.88E-04 & 1.89E-04 & 16.83 & 2.11 & 2.12 \\* \bottomrule
\end{longtable}
\footnotesize{The RV amplitude, K, is given with the upper and lower errors. The eccentricity, eccen, is also given with the upper and lower errors. The orbital period and planet mass ($\sin{i}$) are also given with their upper and lower errors.}

\end{tiny}

\end{landscape}

\onecolumn
\section{Stellar IDs observation summary}

\begin{table}[h!]
\centering
\caption{The different identifiers of all stars in our sample.}
\label{name_conversions}
\begin{tabular}{@{}lllll@{}}
\toprule
NASA Host Name & TIC ID & Gaia DR2 ID & Simbad ID & $\#$RVs\\ \midrule
HD 136352 & 136916387 & Gaia DR2 5902750168276592256 & * nu.02 Lup & 674 \\
HR 858 & 178155732 & Gaia DR2 5064574720469473792 & HD  17926 & 75 \\
GJ 143 & 279741379 & Gaia DR2 4673947174316727040 & HD  21749 & 58 \\
HD 183579 & 320004517 & Gaia DR2 6641996571978861440 & HD 183579 & 71 \\
HD 106315 & 56815340 & Gaia DR2 3698307419878650240 & HD 106315 & 92 \\
HD 18599 & 207141131 & Gaia DR2 4728513943538448512 & HD  18599 & 106 \\
HD 15337 & 120896927 & Gaia DR2 5068777809824976256 & HD  15337 & 118 \\
TOI-431 & 31374837 & Gaia DR2 2908664557091200768 & CD-26  2288 & 174 \\
HD 108236 & 260647166 & Gaia DR2 6125644402384918784 & HD 108236 & 157 \\
HD 73583 & 101011575 & Gaia DR2 5746824674801810816 & HD  73583 & 98 \\
TOI-421 & 94986319 & Gaia DR2 2984582227215748864 & BD-14  1137 & 103 \\
TOI-836 & 440887364 & Gaia DR2 6230733559097425152 & CD-23 12010 & 53 \\
GJ 367 & 34068865 & Gaia DR2 5412250540681250560 & CD-45  5378 & 398 \\
HD 137496 & 346250894 & Gaia DR2 6258810550587404672 & HD 137496 & 142 \\
HD 110113 & 73228647 & Gaia DR2 6133384959942131968 & HD 110113 & 115 \\
TOI-1062 & 299799658 & Gaia DR2 4632865331094140928 & CD-78    83 & 87 \\
TOI-763 & 178819686 & Gaia DR2 6140553127216043648 & CD-39  7945 & 77 \\
TOI-220 & 150098860 & Gaia DR2 5481210874877547904 & CD-61  1276 & 99 \\
TOI-500 & 134200185 & Gaia DR2 5509620021956148736 & CD-47  2804 & 198 \\
TOI-544 & 50618703 & Gaia DR2 3220926542276901888 & HD 290498 & 70 \\
K2-233 & 428820090 & Gaia DR2 6253186686054822784 & BD-19  4086 & 126 \\
GJ 357 & 413248763 & Gaia DR2 5664814198431308288 & L  678-39 & 49 \\
K2-229 & 98720809 & Gaia DR2 3583630929786305280 & BD-05  3504 & 120 \\
TOI-125 & 52368076 & Gaia DR2 4698692744355471616 & TOI-125 & 124 \\
K2-265 & 146364192 & Gaia DR2 2597119620985658496 & BD-15  6276 & 149 \\
GJ 3090 & 262530407 & Gaia DR2 4933912198893332224 & CD-47   399 & 57 \\
TOI-776 & 306996324 & Gaia DR2 3460438662009633408 & LP  961-53 & 64 \\
EPIC 249893012 & 432254760 & Gaia DR2 6259263137059042048 & K2-314 & 77 \\
TOI-1130 & 254113311 & Gaia DR2 6715688452614516736 & TOI-1130 & 76 \\
L 98-59 & 307210830 & Gaia DR2 5271055243163629056 & L   98-59 & 158 \\
LHS 1815 & 260004324 & Gaia DR2 5500061456275483776 & L  181-1 & 72 \\
K2-3 & 173103335 & Gaia DR2 3796690380302214272 & K2-3 & 110 \\
K2-138 & 4610830 & Gaia DR2 2413596935442139520 & K2-138 & 204 \\
K2-32 & 437444661 & Gaia DR2 4130539180358512768 & K2-32 & 245 \\
TOI-270 & 259377017 & Gaia DR2 4781196115469953024 & L  231-32 & 50 \\
TOI-700 & 150428135 & Gaia DR2 5284517766615492736 & TOI-700 & 61 \\
K2-18 & 388804061 & Gaia DR2 3910747531814692736 & K2-18 & 99 \\
LHS 1140 & 92226327 & Gaia DR2 2371032916186181760 & G 268-38 & 291 \\
TOI-269 & 220479565 & Gaia DR2 4770828304936109056 & TOI-269 & 65 \\
GJ 1214 & 467929202 & Gaia DR2 4393265392167891712 & G 139-21 & 165 \\
HD 3167 & 318707987 & Gaia DR2 2554032474712538880 & HD   3167 & 50 \\
K2-266 & 374180079 & Gaia DR2 3855246074629979264 & K2-266 & 63 \\
GJ 1132 & 101955023 & Gaia DR2 5413438219396893568 & L  320-124 & 122 \\
AU Mic & 441420236 & Gaia DR2 6794047652729201024 & HD 197481 & 153 \\
HD 39091 & 261136679 & Gaia DR2 4623036865373793408 & * pi. Men & 555 \\
CoRoT-7 & 280304863 & Gaia DR2 3107267177757848576 & CoRoT-7 & 173 \\
HIP 41378 & 366443426 & Gaia DR2 600698184764497664 & BD+10  1799 & 362 \\
TOI-1052 & 317060587 & Gaia DR2 6357524189130820992 & HD 212729 & 53 \\
HIP 94235 & 464646604 & Gaia DR2 6632318361397624960 & HD 178085 & 58 \\
\bottomrule
\end{tabular}
\end{table}
\footnotesize{ We include the default Host Name from the NASA Exoplanet Archive, the TESS Input Catalogue ID, the Gaia Data Release 2 ID, and the Simbad ID. The final column shows the number of RV observations used in the modelling for each system.}

\end{document}